\documentclass[journal=nalefd,manuscript=letter]{achemso}
\usepackage[utf8]{inputenc}

\usepackage{graphicx,epsfig,amsfonts}
\usepackage{bm}
\usepackage{amsmath}
\usepackage{amssymb}
\usepackage{float}
\usepackage{xcolor}
\usepackage[colorlinks=true,citecolor=blue]{hyperref} 
\usepackage[version=4]{mhchem} 

\usepackage{xcolor}

\graphicspath{{images/}}

\author{Andrey Rybakov}
\affiliation{Instituto de Ciencia Molecular (ICMol), Universitat de València, Paterna 46980, Spain}

\author{Adolfo O. Fumega}
\affiliation{Department of Applied Physics, Aalto University, 02150 Espoo, Finland}

\author{Dorye L. Esteras}
\affiliation{Instituto de Ciencia Molecular (ICMol), Universitat de València, Paterna 46980, Spain}

\author{Jose L. Lado}
\affiliation{Department of Applied Physics, Aalto University, 02150 Espoo, Finland}
\email{jose.lado@aalto.fi}

\author{Jos\'e J. Baldov\'i}
\affiliation{Instituto de Ciencia Molecular (ICMol), Universitat de València, Paterna 46980, Spain}
\email{j.jaime.baldovi@uv.es}

\title{Electrical control of magnons in multiferroic NiI$_2$}

\begin{document}

\begin{abstract}
Layered van der Waals two-dimensional (2D) magnets are a cornerstone
of ultrathin spintronic and magnonic devices.
The recent discovery of a 2D multiferroic with strong magnetoelectric coupling in NiI$_2$
offers a promising platform for the electrical control of spin-wave transport. 
In this work, using \emph{ab initio}
calculations, we investigate how the magnonic properties of monolayer NiI$_2$
can be controlled using an external electric field. We show that 
the emergence of a ferroelectric polarization leads to an energy splitting in the magnon spectrum, thus establishing a way to detect the electric polarization experimentally.
We also show the modulation of the magnon splitting
and the energy position of the singularities in magnon DOS by an electric field due to the strong magnetoelectric coupling.
Our results highlight the interplay between ferroelectricity and magnons in van der Waals
multiferroics
and pave the way to design electrically tunable magnetic devices at the 2D limit.
\end{abstract}

\maketitle

\section{Introduction}

Magnetism in two-dimensional (2D) materials provides a unique
playground for new fundamental phenomena and novel
functional devices\cite{Wang2022,Gibertini2019, Burch2018,Blei_2021}.
Magnetic 2D materials allow modification of their magnetic ground state and magnetic excited states by applying mechanical strain \cite{Cenker2022, Esteras2022}, creating van der Waals heterostructures \cite{Zhang2022}, electrostatic doping \cite{Jiang2018, Huang2018, Zhang2020}, intercalation \cite{Mishra2024, Iturriaga2023} and molecular deposition \cite{Ruiz2024}.
These systems range from conventional ferromagnets\cite{Huang2017,BedoyaPinto2021},
quantum spin liquids\cite{Ruan2021} and heavy-fermion Kondo insulators\cite{Vao2021,Posey2024}
dominated by quantum fluctuations, and recently
multiferroic 2D materials\cite{Song2022,Amini2024,Gao2024} featuring entangled magnetism and ferroelectricity.
Importantly, 2D materials are prominent as building-blocks for the realization of the spin-wave circuits at the limit of miniaturization, where spin waves (SW) are utilized as an information carriers in contrast to conventional electronics \cite{Barman2021, Flebus2024}.

Multiferroic materials offer promising building-blocks to achieve electrical control over magnon transport.
This class of materials can display simultaneously ferroelectric and magnetic order and they exhibit a strong magnetoelectric coupling between the two, which makes them highly attractive from a technological point of view \cite{Kimura2003, Hur2004, Fiebig2005, Gajek2007, Pantel2012, Hu2015, Matsubara2015}.
Special attention has been recently given to van der Waals multiferroics due to the enlarged possibilities that these compounds offer: a way to engineer novel heterostructures with emergent functionalities \cite{Fumega2022, Fumega2023, Amini2024}.
In this realm, NiI$_2$ is the paradigmatic van der Waals multiferroic and is attracting a growing attention \cite{Billerey1977, Kuindersma1981, Kurumaji2013, Liu2020, Ju2021, Song2022, Amini2024}.
Its multiferroic order is a consequence of non-collinear magnetic spin-spiral and the strong spin-orbit coupling provided by the iodine atoms \cite{Fumega2022}.
Recently, it has been demonstrated that the multiferroic order of NiI$_2$ can survive in the few-layer and single-layer limits \cite{Ju2021, Song2022, Amini2024}.
Therefore, establishing monolayer NiI$_2$ as the first purely 2D multiferroic and adding a new system to the family of van der Waals compounds to engineer artificial materials \cite{Geim2013}.
However, a theoretical investigation of the tunability of the magnon spectrum of multiferroic NiI$_2$ with external electric fields has remained unexplored.

In this work, combining effective models and \emph{ab initio} calculations, 
we show how the magnon spectrum of monolayer NiI$_2$ can be controlled with an external electric field.  
We observe that the C$_6$ rotation symmetry breaking associated with the emergence of ferroelectricity is directly manifested as a splitting $\Theta$ in the magnon spectrum occurring at the rotational image of the $\mathbf{q}$-vector of the spin spiral.
We show that the size of the magnon splitting can be controlled with an external electric field.
Our results highlight a way to experimentally estimate the magnitude of the electric dipole through the splitting
in the magnon spectrum and demonstrate that multiferroic NiI$_2$ is a promising platform for electrically tunable magnon transport.

\begin{figure*}
    \centering
    \includegraphics[width=\textwidth]{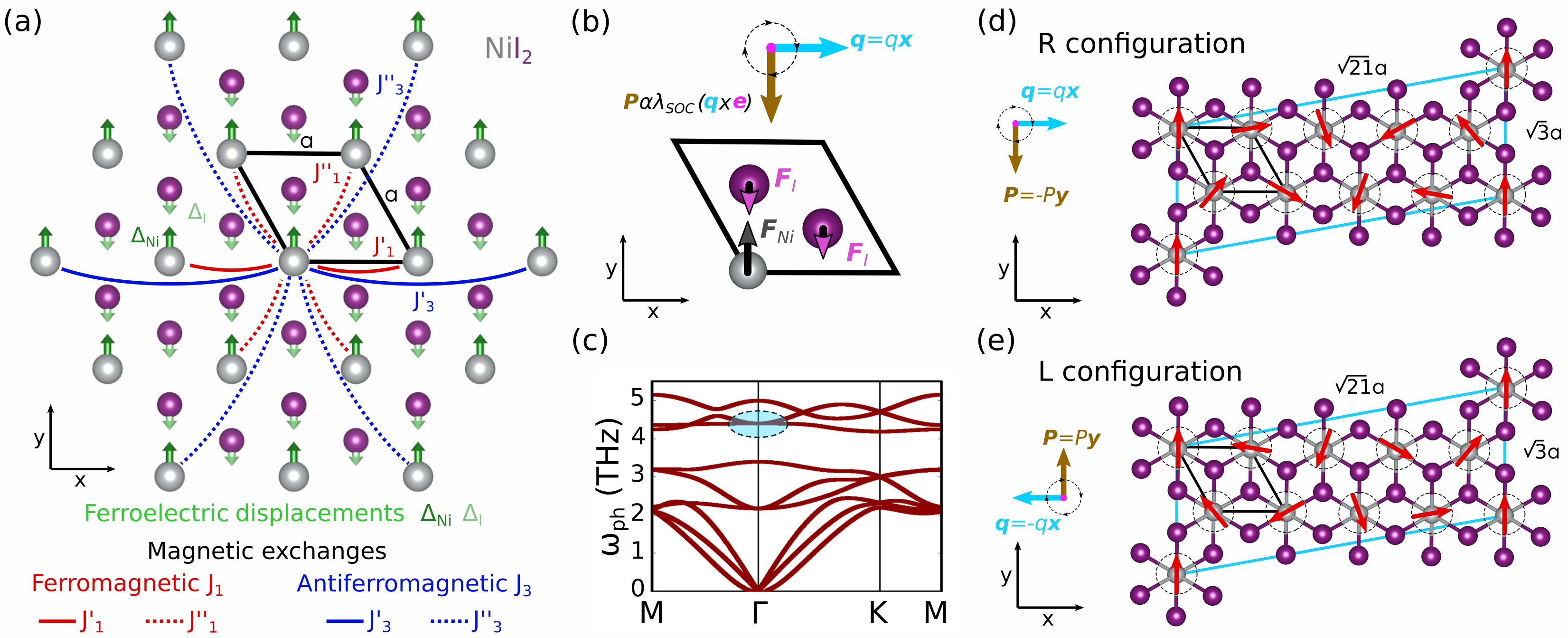}
    \caption{
        (a) Schematic of the model used to study the magnons in multiferroic NiI$_2$.
        The structural unit cell is shown in black with lattice parameter $a$ and Ni (I) atoms in gray (purple).
        The ferroelectric displacements on the atoms are depicted as green arrows, in the direction where the average electric polarization emerges. The first $J_1$ and third $J_3$ neighbor magnetic exchange couplings are shown in red and blue respectively.
        The ferroelectric distortion breaks the crystal's C$_6$ symmetry, leading to two non-equivalent $J^{\prime}$ and $J^{\prime\prime}$ magnetic exchanges for both -- first and third -- groups of neighbors.
        (b) Schematic of the emergent ferroelectric polarization perpendicular to the plane formed by the propagation vector $\mathbf{q}$ and the rotation vector $\mathbf{e}$ of the spin spiral.
        The ferroelectric forces associated with the emergent polarization are directed parallel to it.
        (c) Phonon band structure for monolayer NiI$_2$. The optical phonon mode associated with the ferroelectric displacements $\omega_{FE}$ is highlighted in blue.
        (d,e) Equivalent spin spiral configurations with opposite emergent electric polarization used to calculate the ferroelectric forces. Commensurate $\sqrt{3}a\times\sqrt{21}a$ supercells are used to describe the spin spiral associated with the $J_3/J_1$ ratio predicted by DFT calculations.
        The red arrows correspond to the spins.
    }
    \label{Fig:model_scheme}
\end{figure*}

\section{Results and discussion}

\subsection{Magnons in multiferroic \ce{NiI_2}}

\begin{figure}[ht!]
    \centering
    \includegraphics[width=\textwidth]{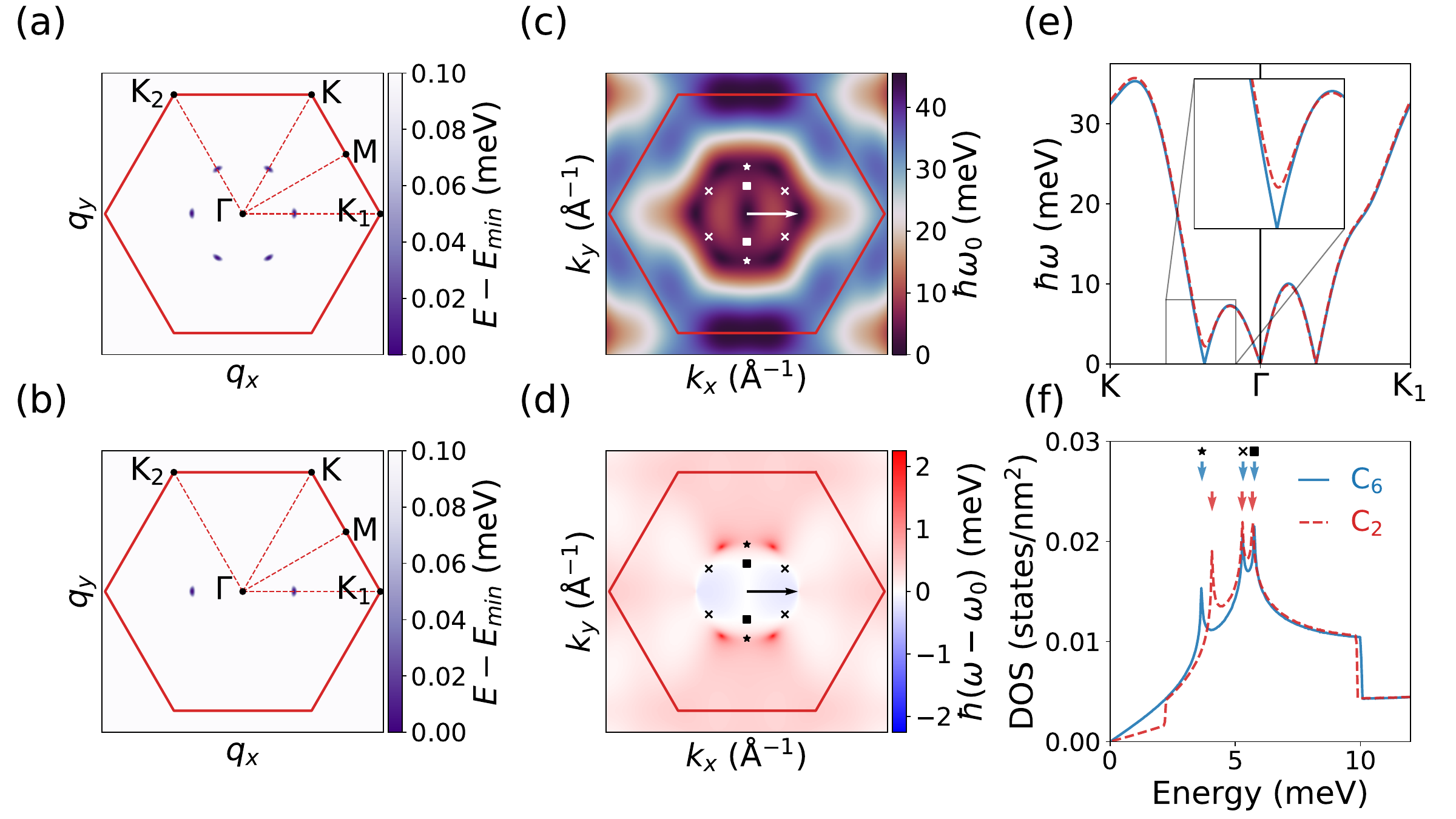}
    \caption{
        (a) Ground state energy of~\eqref{eq:model_hamiltonian} under variation of the spin-spiral vector $\mathbf{q}$ for undistorted structure with C$_6$ symmetry;
        (b) Same as (a), but for distorted structure with C$_2$ symmetry;
        (c) Magnon dispersion for the first Brillouin zone of the undistorted structure;
        (d) Modification of magnon energies via the introduction of the ferroelectric order;
        (e) Magnon energies for the undistorted (blue, solid) and distorted (red, dashed) structures. Inset shows the splitting;
        (f) Magnon density of states (DOS) for the undistorted (blue, solid) and distorted (red, dashed) structures.
        Arrows show positions of the peaks for the first three van Hove singularities;
        In plots (a-d) first Brillouin zone is indicated as a red hexagon.
        In plots (c,d) ordering vectors of spin-spiral $\mathbf{q}$ are plotted as white or black arrows, positions of saddle points that correspond to the first three van Hove singularities in DOS of plot (f) are marked with white or black markers.
    }
    \label{Fig:magnons_multiferroic}
\end{figure}

We start by addressing the magnon spectrum in monolayer NiI$_2$.
The structural unit cell of monolayer NiI$_2$ is shown in Fig.~\ref{Fig:model_scheme}a.
The unpaired $d$ electrons of Ni atoms lead to a spin state $S=1$ with ferromagnetic ($J_1$) and antiferromagnetic ($J_3$) couplings between first and third neighbors respectively.
Due to the competition between the exchange couplings, monolayer NiI$_2$ develops a spin-spiral order at low temperatures.
The spin spiral is characterized by a propagation vector $\mathbf{q}$, that determines its periodicity and direction, and rotation vector $\mathbf{e}$, that determines the rotation plane of the spiral.
In the case of a spin spiral dominated by $J_3/J_1$ ratio, the propagation vector in the reciprocal space lies along the $\Gamma$K segment (see Fig.~\ref{Fig:magnons_multiferroic}a for definition of high symmetry points), corresponding to a propagation vector in real space along the third neighbors' direction (Fig.~S1). 
In bulk NiI$_2$, it has been shown that interlayer and Kitaev interactions can lead to an out-of-plane component of the spin spiral \cite{Li2023, Kapeghian2024}.
However, in the monolayer limit, the magnetic order is dominated by the $J_3/J_1$ ratio \cite{Amini2024, Sdequist2023}.
Therefore, its magnetic order can be described with the following Heisenberg Hamiltonian 

\begin{equation}\label{eq:model_hamiltonian}
    \mathcal{H} = \sum_{i j } \mathcal{J}_{ij} \mathbf{S}_{i} \cdot \mathbf{S}_{j} + 
    \mathcal{H}^{\text{ani}}
\end{equation}
where the most important exchange couplings correspond to first $J_1$
and third neighbors $J_3$ (see SI for details).
The term $\mathcal{H}^{\text{ani}}$ corresponds to the anisotropic exchange
couplings arising as a result of
spin-orbit coupling, which have a minimal effect on the spin-spiral q-vector.

Due to the strong spin-orbit coupling ($\lambda_{SOC}$) originating from the I atoms \cite{Fumega2022}, NiI$_2$ develops an electric polarization $\mathbf{P}$ in the direction perpendicular to the propagation and rotation vectors of the spin spiral \cite{Mostovoy2006}.
Note that $\mathbf{P}$ corresponds to the average emergent polarization.
Experimentally it has been demonstrated that a modulated polarization with half of the periodicity of the spin spiral emerges, allowing the atomic scale visualization of the multiferroic order \cite{Amini2024}.
For simplicity, we will consider the average emergent polarization, although the model could be extended to consider the effect of the ferroelectric modulation.
The emergence of the ferroelectric order is accompanied by emanating ferroelectric forces ($\mathbf{F}_{Ni}$ and $\mathbf{F}_{I}$ in Fig. \ref{Fig:model_scheme}b) parallel to the polarization's direction causing a ferroelectric displacement $\boldsymbol{\Delta}_{Ni}$ and $\boldsymbol{\Delta}_{I}$ in each of the atoms (as shown in the schematic of Fig. \ref{Fig:model_scheme}a).
All the parameters entering this process can be computed \emph{ab initio}. 

We start estimating the values of $J_1$ and $J_3$ in the undistorted structure using density functional theory (DFT) \cite{Giannozzi2017, Giannozzi2009, Dudarev1998, Ernzerhof1999, Hamann2013, Monkhorst1976, Timrov2018, Timrov2022, Pizzi2020, He2021, Solovyev2021} (see SI for computational details).
Obtained values allow us to determine the $\mathbf{q}$ vector of the spin spiral via minimization of the energy within the first Brillouin zone

\begin{equation}\label{eq:energy_qspiral}
    E(\mathbf{q}) \propto \sum_{ij}\mathcal{J}_{ij}\cos{\left(\mathbf{q}\cdot\mathbf{r}_{ij}\right)}
\end{equation}
where  $\boldsymbol{r}_{ij}$ is the vector between two corresponding magnetic centers.
Spin vectors are considered to be in the $xy$ plane and the rotational axis is assumed along the direction of the $z$ axis, due to the isotropic nature of the Hamiltonian (eq. \eqref{eq:model_hamiltonian}). 
Due to the $C_6$ symmetry of the undistorted structure, $6$ degenerate energy minima are exhibited in the first Brillouin zone (Fig. \ref{Fig:magnons_multiferroic}a).
With $J_1 = -2.8$~meV and $J_3 = 2.32$~meV we obtained $\mathbf{q}=(0.124,0.124,0)$ (written in the basis of reciprocal vectors) along the $\Gamma$K path, which is in good agreement with recent theoretical \cite{Sdequist2023} and experimental \cite{Amini2024} studies.
The approximate commensurate spin-spiral associated with the predicted $\mathbf{q}$ is a $\sqrt{3}a\times\sqrt{21}a$ (Fig. \ref{Fig:model_scheme}d).
When the spin-orbit coupling is included in the DFT calculations, the ferroelectricity emerges indicated by the ferroelectric forces 
$\mathbf{F}_{Ni}$ and $\mathbf{F}_{I}$ (Fig. \ref{Fig:model_scheme}b) that induce the ferroelectric displacements  $\boldsymbol{\Delta}_{Ni}$ and $\boldsymbol{\Delta}_{I}$ in each atom.
The ferroelectric forces can be computed from DFT by considering two equivalent spin-spiral configurations (Figs. \ref{Fig:model_scheme}d and \ref{Fig:model_scheme}e) with opposite $\mathbf{q}$ vectors\cite{Fumega2022}.
The difference between the emergent forces in each configuration allows us to factor out other contributions to get $\mathbf{F}_{Ni}=(0,-8.8,0)$~meV/\AA\  and $\mathbf{F}_{I}=(0,4.4,0)$~meV/\AA\ .
We can estimate the ferroelectric displacements in the absence of an electric field from the ferroelectric forces as

\begin{equation}\label{eq:ferroelectricdisp}
    \boldsymbol{\Delta}_{\alpha} = \frac{\mathbf{F}_{\alpha}}{k_{\alpha}},
\end{equation}
where $k_{\alpha}$ are the elastic constants for each species $\alpha = (\mathrm{Ni}, \mathrm{I})$.
The elastic constants can also be computed from DFT by calculating the harmonic phonon spectrum (Fig.~\ref{Fig:model_scheme}c) and identifying the frequency of the phonon mode associated with the ferroelectric displacements $\omega_{FE}=4.4$ THz.
Taking into account the atomic masses of each species $m_{\alpha}$ the elastic constants are given by 

\begin{equation}\label{eq:elasticconstants}
    k_{\alpha} = m_{\alpha}\omega_{FE}^2.
\end{equation}

We obtain $k_{Ni}=146$ meV/\AA$^2$, $k_{I}=315$ meV/\AA$^2$, which using eq. (\ref{eq:ferroelectricdisp}) lead to $\Delta_{Ni}=0.06$~\AA\  and $\Delta_{I}=0.014$~\AA\   in the opposite direction (Fig.~\ref{Fig:model_scheme}a).
An effective ferroelectric dipole $p=de^*=0.003$~e\AA\  can be estimated by considering an average atomic displacement from the computed ones $d=0.03$~\AA\ and an effective charge $e^*=\beta e$, with $e$ the charge of the electron and $\beta=0.1$ a constant that can be extracted from the Born effective charges.

The introduction of the ferroelectric distortion
breaks the C$_6$ symmetry of the crystal, leading to a C$_2$ symmetry.
As a result, the six first ($J_1$) and third ($J_3$) nearest neighbor exchange interactions in the spin model \eqref{eq:model_hamiltonian} break their degeneracy resulting in the new sets of \{$J_1^{\prime}$, $J_1^{\prime\prime}$\} and \{$J_3^{\prime}$, $J_3^{\prime\prime}$\} magnetic exchanges respectively as shown in Fig.~\ref{Fig:model_scheme}a.
We can compute for this distorted structure the new set of magnetic exchange interactions (see SI for the values) and get the ground state $\mathbf{q}$ vector that minimizes the energy using eq.~\eqref{eq:energy_qspiral}.
Due to the symmetry breaking, there are only two energy minima degenerated (Fig. \ref{Fig:magnons_multiferroic}b) in the distorted structure in contrast to the six degenerate minima (Fig. \ref{Fig:magnons_multiferroic}a) occurring in the undistorted structure.

\subsection{Magnons in multiferroic \ce{NiI_2}}
Magnon excitations appear as fluctuations of the spin spiral magnetic order of NiI$_2$.
The magnetic ground state for the undistorted and the ferroelectric structures allows
us to define the spin-wave excitations.
Due to the magnetoelectric coupling of this material, 
some magnon branches will couple to the ferroelectric order,
giving rise to electromagnon excitations\cite{PhysRevB.108.064414,Gao2024}.
The magnon Hamiltonian is obtained via the expansion of the Heisenberg Hamiltonian~\eqref{eq:model_hamiltonian} with a Holstein-Primakoff transformation \cite{Holstein1940} within the linear approximation \cite{Toth2015} and consecutive diagonalization of bosonic Hamiltonian \cite{White1965}.
For each spin of the lattice the quantization axis is chosen along the direction defined by the spiral
configuration and the transversal excitations are defined in the plane perpendicular to the spin direction.
We arrive at the expression for the magnon Hamiltonian (see SI for details), whose eigenfrequencies take the form\cite{Cong2024}
$   \mathcal{H}^{LSWT}
    =
    \sum_k
    \hbar\omega(\boldsymbol{k})
    \biggl[
      \beta_k^{\dagger}\beta_k + \dfrac{1}{2}
    \biggr]
$, with 
$
    \hbar\omega(\mathbf{k}) = 2S\cdot\sqrt{A^2(\boldsymbol{k}) - B^2(\boldsymbol{k})}
$, 
$
    A(\boldsymbol{k}) = \sum_j J_j \biggl(\bigl[1 + \cos(\mathbf{q}\cdot\boldsymbol{\delta}_j)\bigr]\cdot\cos(\mathbf{k}\cdot\boldsymbol{\delta}_j) - 2\cos(\mathbf{q}\cdot\boldsymbol{\delta}_j)\biggr)
$, 
$
    B(\boldsymbol{k}) = \sum_j J_j \bigl[1 - \cos(\mathbf{q}\cdot\boldsymbol{\delta}_j)\bigr]\cdot\cos(\mathbf{k}\cdot\boldsymbol{\delta}_j)
$
where $\boldsymbol{\delta}_j$ are the vectors connecting first and third nearest neighbors and index $j$ runs over the set of pairs \{$(J_1^{\prime}, \boldsymbol{\delta}_1)$, $(J_1^{\prime\prime}, \boldsymbol{\delta}_2)$, $(J_1^{\prime\prime}, \boldsymbol{\delta}_3)$, $(J_3^{\prime}, 2\boldsymbol{\delta}_1)$, $(J_3^{\prime\prime}, 2\boldsymbol{\delta}_2)$, $(J_3^{\prime\prime}, 2\boldsymbol{\delta}_3)$\}.

The magnon dispersion of the undistorted structure does not exhibit C$_6$ symmetry as it depends on the choice
between three equivalent spiral vectors: along $\Gamma$K$_1$, $\Gamma$K and $\Gamma$K$_2$.
However, Brillouin zone-resolved magnon energies for three spiral vectors are equivalent upon rotation by $0^{\circ}$, $-60^{\circ}$, $-120^{\circ}$ correspondingly (see Fig.~S2).
In Fig.~\ref{Fig:magnons_multiferroic}c we present the magnon dispersion for the undistorted structure with the $\mathbf{q}$-vector along $\Gamma$K$_1$ direction, for direct comparison with magnon energies of the structure with emergent polarization.
The introduction of the spiral keeps the C$_6$ symmetry of the points with minimal energies, but the excitations with $\mathbf{k}$ vectors along the paths $\Gamma$K and $\Gamma$K$_1$ are not equivalent in general (see Fig.~\ref{Fig:magnons_multiferroic}e). 
Inclusion of the emerging polarization introduces non-uniform shift of the
magnon energies across the whole Brillouin zone (see Fig.~\ref{Fig:magnons_multiferroic}d).
However, the major effect is the opening of the splitting due to the breaking of C$_6$ rotation symmetry.
The appearance of a splitting in one of the magnon branches in both cases can be accounted as characteristic
signature of the emerging ferroelectric order in \ce{NiI_2}.

The spectra of magnonic excitations can be characterized by the
magnon density of states (DOS).
There are $6$ van Hove singularities \cite{VanHove1953}, that correspond
to the six groups of energy degenerate saddle points in the magnon dispersion.
We plot only three singularities in the low energy region (Fig.~\ref{Fig:magnons_multiferroic}f)  that correspond to
the saddle points depicted in Figs.~\ref{Fig:magnons_multiferroic}c,d as only
those points can be observed at low temperatures (see SI for the full energy range).
Interestingly, the peak that corresponds to the saddle point is defined by the intersections of two lines.
First, the line perpendicular to the spin-spiral vector and second, the
line connecting two energy minima where the splitting is opened upon ferroelectric distortion (stars in Fig.~\ref{Fig:magnons_multiferroic}).
This second line changes as the distortion gets stronger,.
The magnon state, that would correspond to this point in momentum space,
can be describe as a set of one-dimensional spirals along the $x$ axis
with the phase shift of magnetic excitations between them and each fifth spiral being in phase.
Such magnetic excitations can be probed via inelastic
tunneling spectroscopy\cite{Heinrich2004,Klein2018}, 
or electrically driven paramagnetic resonance \cite{Baumann2015,2024arXiv240314145W},
allowing to directly image the magnon DOS
at the atomic scale\cite{Spinelli2014,Ganguli2023}.

\begin{figure}[ht!]
     \centering
     \includegraphics[width=\textwidth]{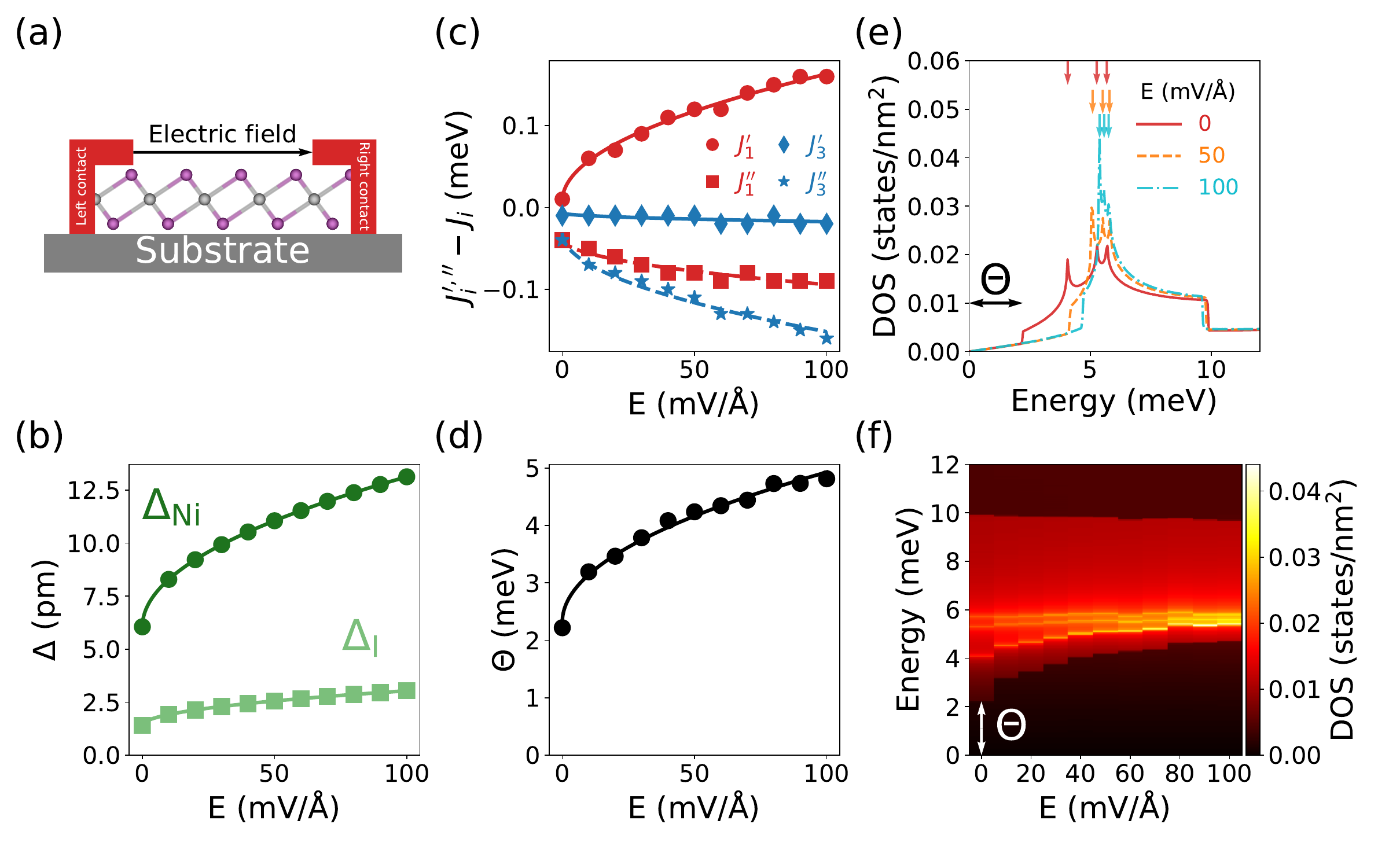}
    \caption{
        (a) Sketch of the possible device for the application of electric field, gray (purple) circles depict \ce{Ni} (\ce{I}) atoms.
        (b) Ferroelectric displacements vs applied electric field. Values that were used in simulations are plotted with circles (squares) for \ce{Ni} (\ce{I}), lines are the square root fits ($\Delta \propto \sqrt{E}$). Zero corresponds to the undistorted structure;
        (c) Magnetic exchange parameters vs applied electric fields. Values are relative to the ones of undisturbed structure;
        (d) Splitting in the magnon energies at the rotational image of the spin-spiral vector $\mathbf{q}$ along the $\Gamma$K path vs electric field. Circles are for the calculated values, line is a square root fit (Gap $\propto \sqrt{E}$);
        (e) Magnon DOS as for three values of applied electric field: $0$~mV/\AA\  (red, solid), $50$~mV/\AA\  (orange, dashed), $100$~mV/\AA\  (cyan, dash-dotted).
        Arrows indicate positions of the first three van Hove singularities;
        (f) Magnon DOS for the full range of energies and applied values of electric field. White line separates DOS of the C$_2$ symmetric (undistorted) structure.
    }
    \label{Fig:electric_field}
\end{figure}

\subsection{Electrical control of magnons}

Finally, we analyze the control of spin-waves using an external electric field.
The emergence of an electric dipole in multiferroic NiI$_2$ allows coupling magnons to 
an external electric field $\mathbf{E}$. The conceptual scheme of the device for the measurement at
the monolayer limit was demonstrated recently \cite{Higashitarumizu2020} and depicted in Fig.~\ref{Fig:electric_field}a.
This phenomena can be effectively modeled in a linear response regime, where
the energy of the electric dipole is balanced by the energy of the harmonic ferroelectric displacements
$
    \mathbf{p}\cdot \mathbf{E} = \sum_{\alpha}\dfrac{k_{\alpha}\Delta_{\alpha}^2}{2}.
$. Note that only the component of the electric field that is parallel to the electric polarization couples to it
in lowest order. Therefore, we will consider an external field parallel to the polarization.
Considering the relationship between the ferroelectric displacements of the two species from the force balance as $\Delta_{Ni}=2k_{I}/k_{Ni}\Delta_{I}$, we can estimate the contribution to the two ferroelectric displacements introduced by the external electric field ($\Delta_{Ni}(E)$ and $\Delta_{I}(E)$) as
$
    \Delta_{I}(E)= \sqrt{\dfrac{k_{Ni}}{(2k_{I}+k_{Ni})k_{I}}Ep}.
$.
The total atomic displacements are the sum of the displacements caused by the
emergent ferroelectric dipole and the distortion caused by the external electric field.
Figure \ref{Fig:electric_field}b shows the ferroelectric displacements as a function of an external electric field.
Effectively, the distortions induced by the electric field modify the set of exchange parameters \{$J_1^{\prime}$,$J_1^{\prime\prime}$\} and \{$J_3^{\prime}$,$J_3^{\prime\prime}$\}, keeping the symmetry of magnetic exchange interactions of the distorted structure as shown in Fig. \ref{Fig:electric_field}c.
As $\sqrt{E} \propto \Delta$ and changes of the exchange couplings are proportional to $\sqrt{E}$ as well, we conclude that applied values of electric field induce linear response of the exchange couplings $J \propto \Delta$.
With the new set of exchange parameters we apply the same analysis of the magnetic excitations as in the previous section.
In Fig.~\ref{Fig:electric_field}d the evolution of the splitting in the magnon energies follows a similar trend as
the exchange couplings, as evident from the equation \eqref{eq:model_hamiltonian} (see SI
for the changes of the magnon dispersion over the first Brillouin zone).
In the same way as for the splitting caused by the emerging ferroelectric order, the dependency of the splitting height
with the electric field can be probed using dynamical measurements of the magnetic response\cite{Gao2024},
or magnetic resonance\cite{Qin2021,PhysRevLett.125.147201}.

Following the fact that the electric field causes distortion that is aligned with the distortion caused by the symmetry breaking, the response of the first van Hove singularity to the applied electric field is more robust.
In Figs.~\ref{Fig:electric_field}e-f we plot the magnon DOS for the low energy region.
One can see that the lower energy peak approaches the second and third, while the latter once show weak sensitivity to the electric field.
That behavior opens a way for the non-invasive selective control of magnon
van Hove singularities in magnetic system. 
This electric control of magnons should be contrasted with the case
of electronic systems, where the control of 
van Hove singularities usually requires twisting of two layers \cite{Li2009}, chemical doping \cite{Markiewicz1990} or gating \cite{Novoselov2004}.
The application of electric field in multiferroic material offers a controllable and
reversible knob to engineer the magnonic dispersion.

\section{Conclusions}
The electrical control of magnon transport is one of milestones in ultrathin spintronic devices.
Here, we demonstrated a mechanism to control the
magnon dispersion in the monolayer multiferroic NiI$_2$
by leveraging its intrinsic magnetoelectric coupling.
Our mechanism exploits that ferroelectric order in NiI$_2$ strongly couples
to the magnon spectra, providing a strategy to modify spin waves transport with external electric fields. 
Our strategy leverages the microscopic mechanism leading to multiferroicity in NiI$_2$,
namely the combination of the spin-spiral order and strong spin-orbit coupling,
which leads to displacements on the atomic positions. 
We first showed that the atomic displacements associated to the ferroelectricity
lead to a modification in the magnetic exchange interactions,
directly impacting the magnon spectra.
Thanks to the emergence of in-plane ferroelectric order, an in-plane bias allows to directly
modify the structural distortion by directly coupling to the ferroelectric polarization.
We showed that such electric control of the magnon dispersion enables
controlling the low energy dispersion of the magnon modes. In particular,
a single-layer of NiI$_2$ features a magnon branch whose splitting
is electrically controlled by the in-plane bias.
Ultimately, our mechanism establishes a strategy 
to control low energy magnon transport electrically, opening a pathway towards
designing ultrathin electrically controllable magnonic devices.

\begin{acknowledgement}
This work was supported by the European Union (ERC-2021-StG-101042680 2D-SMARTiES) and the Excellence Unit “María de Maeztu” (CEX2019-000919-M). J.J.B. and A.R. acknowledge the Generalitat Valenciana (grant CIDEXG/2023/1 and PhD scholarship GRISOLIAP/2021/038, respectively). A.O.F. and J.L.L. acknowledges financial support from
the Academy of Finland Projects Nos. 331342, 358088, and 349696,
the Jane and Aatos Erkko Foundation,
and the Finnish Quantum Flagship. 
A.R. thanks Jaime Ferrer for fruitful discussions. We acknowledge the computational resources provided by the Aalto Science-IT project and  HPC systems Cobra and Raven at the Max Planck Computing and Data Facility.
\end{acknowledgement}

\begin{suppinfo}
Detailed calculation scheme and relevant simulation parameters are available in the Supporting Information.
\end{suppinfo}

\bibliography{biblio}

\end{document}


\newpage

\section{Geometry}

\begin{figure}[H]
    \centering
    \includegraphics[width=0.9\linewidth]{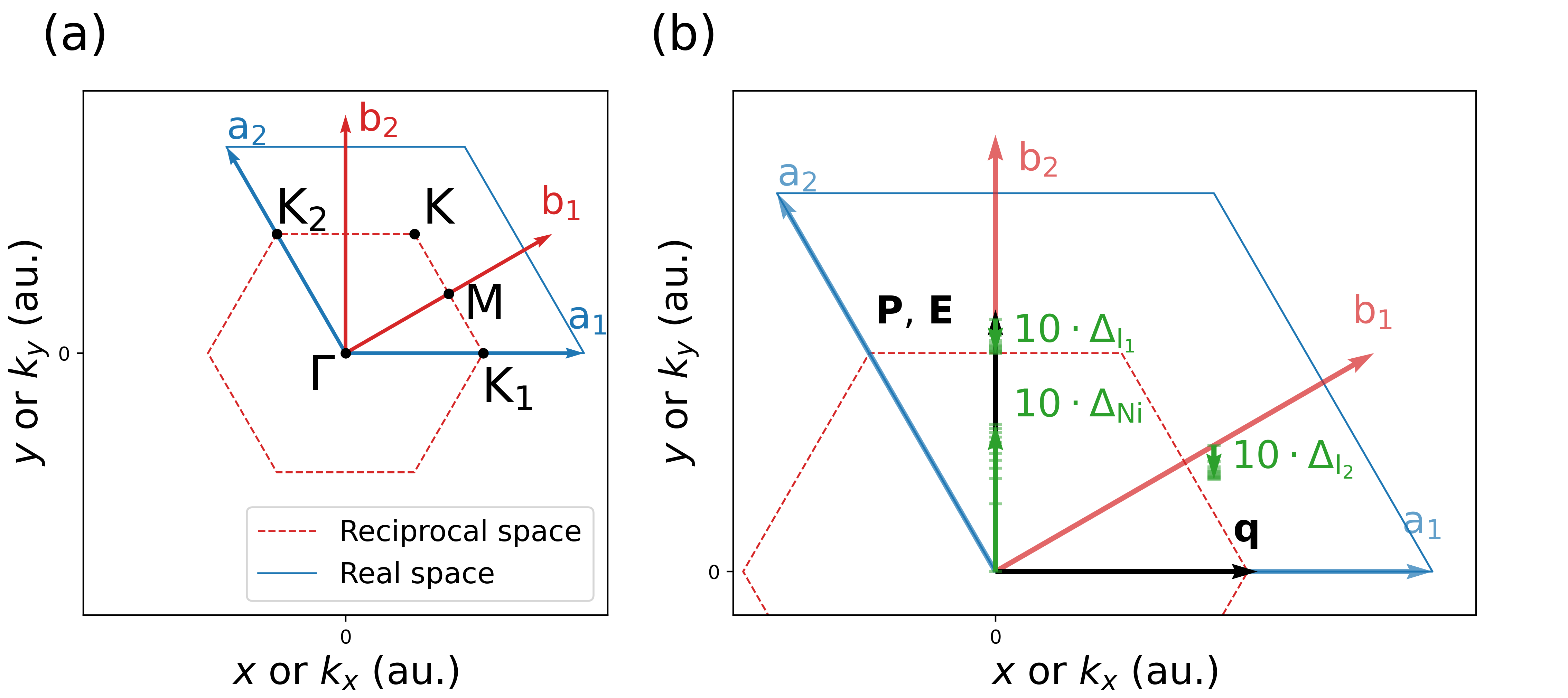}
    \caption{(a) Correspondence between the directions of the real and reciprocal space. First Brillouin zone and reciprocal lattice vector are scale to match the volume of the primitive unit cell for display purposes. Relevant high symmetry k-points are depicted. (b) Structure distortion and relevant vectors of the real space and reciprocal space:
    $\mathbf{b}_1$, $\mathbf{b}_2$ are the reciprocal lattice vectors.
    $\mathbf{a}_1$, $\mathbf{a}_2$ are the lattice vectors.
    $\mathbf{P}$ is the polarization vector.
    $\mathbf{q}$ is the spiral vector of the magnetic ground state.
    $\Delta_{\ce{Ni}}$, $\Delta_{\ce{I_1}}$, $\Delta_{\ce{I_2}}$ are the displacements of the \ce{Ni}, \ce{I_1}, \ce{I_2} atoms, respectively.
    Starting position for the atom displacement corresponds to the undistorted structure.
    Final position corresponds to the structure under the applied electric field of $100$~mV/\AA.
    Atom displacements are scaled with the factor of $10$ for the visualization purposes.
    All vectors indicates the the directions of corresponding quantities. Vectors of atom displacements reflect relative values with respect to the real space.}
    \label{fig:distortion-summary}
\end{figure}

Distortion of the structure introduced by the SOC and applied electric field is summarized in Fig.~\ref{fig:distortion-summary}(b), where the relevant vectors in the real space are displayed as well. In Figure ~\ref{fig:distortion-summary}(a) we display the first Brillouin zone with the high symmetry k-points $\Gamma$, K, M and two rotational images of the point K: $\mathrm{K_1}$ and $\mathrm{K_2}$.

\section{DFT}

In order to obtain the exchange parameters we performed DFT calculations as implemented in the Quantum Espresso (QE) \cite{Giannozzi2009, Giannozzi2017} package. Fully relativistic norm-conserving pseudopotentials \cite{Hamann2013} and Perdew–Burke–Ernzerhof (PBE) functional were used to describe the exchange-correlation energy \cite{Ernzerhof1999}. The description of the electronic wave functions was converged using wave function and charge density cut-offs of 120 and 500 Ry, respectively. In the calculations the Brillouin zone was sampled using a Monkhorst-Pack grid of 8 $\times$ 8 $\times$ 1 \cite{Monkhorst1976}. The strong electron correlation of \ce{NiI_2} was described within the DFT + U formalism, according to the simplified scheme proposed by Dudarev \cite{Dudarev1998}. To determine the optimal Hubbard parameter (U), self-consistent calculations based on the linear response method were carried out \cite{Timrov2018, Timrov2022}. To achieve this, the parameters of the Monkhorst-Pack grid were carefully converged to 3 $\times$ 3 $\times$ 1. Under these conditions, the Hubbard parameter was converged after two self-consistent + liner response iterations to a value of $\sim$ 5 eV, that was used for the rest of the calculations. Full spinor calculations were performed with the magnetisation direction along three principal axes ($xyz$).

\section{Wannier}
Then the set of Kohn-Sham states were projected into the real space onto the basis of localized Wannier orbitals as implemented in Wannier90 package \cite{Pizzi2020}.  Quality of the fit was characterised in a standard way by two parameters:

\begin{equation}
    \eta_{\mathrm{max}} = \max_{n\boldsymbol{k}}(\vert\varepsilon_{n\boldsymbol{k}}^{\mathrm{DFT}} - \varepsilon_{n\boldsymbol{k}}^{\mathrm{Wannier}}\vert) 
\end{equation}
\begin{equation}
    \eta_{\mathrm{average}} = \sqrt{\dfrac{1}{N}\sum_{n\boldsymbol{k}}(\varepsilon_{n\boldsymbol{k}}^{\mathrm{DFT}} -  \varepsilon_{n\boldsymbol{k}}^{\mathrm{Wannier}})^2}
\end{equation}

\section{TB2J}

The set of exchange values was calculated with TB2J package \cite{He2021} considering the basis of both \ce{Ni} and \ce{I} atoms with further downfolding \cite{Solovyev2021} of the model onto the net of only \ce{Ni} atoms. $512$ integration steps were used  in the computation of Green functions, with k-mesh of $10\times10\times1$ and the maximum interatomic distance of $25$ \AA. For the downfolding the q-mesh of $20\times20\times1$ was used. Isotropic exchange was defined as the $\mathrm{Tr}(\mathbf{J})/3$, where $\mathbf{J}$ is a full exchange tensor. The values of the exchange parameters for the high symmetry phase are $J_1 = -2.8$~meV, $J_3 = 2.32$~meV, values for the ferroelectric phase summarised in the Table~\ref{table:parameters}

\begin{table}[H]
    \centering
    \begin{tabular}{|c|c|c|c|c|c|c|c|c|c|c|c|c|}
     \hline
        E, mV/\AA    &     0 &    10 &    20 &    30 &    40 &    50 &    60 &    70 &    80 &    90 &   100 \\ \hline
        $J_1^{\prime}$, meV       & -2.79 & -2.74 & -2.73 & -2.71 & -2.69 & -2.68 & -2.68 & -2.66 & -2.65 & -2.64 & -2.64 \\ \hline
        $J_1^{\prime\prime}$, meV & -2.84 & -2.85 & -2.86 & -2.87 & -2.88 & -2.88 & -2.89 & -2.88 & -2.89 & -2.89 & -2.89 \\ \hline
        $J_3^{\prime}$, meV       &  2.31 &  2.31 &  2.31 &  2.31 &  2.31 &  2.31 &  2.30 &  2.30 &  2.31 &  2.30 &  2.30 \\ \hline
        $J_3^{\prime\prime}$, meV &  2.28 &  2.25 &  2.24 &  2.23 &  2.22 &  2.21 &  2.19 &  2.19 &  2.18 &  2.17 &  2.16 \\ \hline
    \end{tabular}
    \caption{Six isotropic exchange parameters of the effective model for distorted structure ($E = 0$) and for the structures under the applied electrical field.}
    \label{table:parameters}
\end{table}

\section{Spiral order}
Note that the periodicity of the spin spiral predicted by DFT calculations differs from the one found experimentally, as detailed in the reference \cite{Amini2024}.
The disagreement could be related to substrate effects in the experiment or to overestimations of the long-range magnetic exchange interactions for DFT functionals.
We consider the q predicted by DFT for consistency in the theoretical modeling.
However, the qualitative results that we obtained remain unaffected.

\section{Magnon dispersion}

We compute magnon dispersion for the following Hamiltonian:

\begin{equation}\label{Eq:model_hamiltonian}
    \mathcal{H} = \sum_{i} \sum_j
    J_j \mathbf{S}_{i}^T \left(\mathbf{S}_{i+\delta_j} + \mathbf{S}_{i-\delta_j}\right),
\end{equation}
where index $i$ runs over the unit cells of the crystal and index $j$ runs over the set of pairs \{$(J_1^{\prime}, \boldsymbol{\delta}_1)$, $(J_1^{\prime\prime}, \boldsymbol{\delta}_2)$, $(J_1^{\prime\prime}, \boldsymbol{\delta}_3)$, $(J_3^{\prime}, 2\boldsymbol{\delta}_1)$, $(J_3^{\prime\prime}, 2\boldsymbol{\delta}_2)$, $(J_3^{\prime\prime}, 2\boldsymbol{\delta}_3)$\}; $\mathbf{S}_i = (S_i^x, S_i^y, S_i^z)^T$ is a vector. and $J_j$ are isotropic exchange constants.
$J_{1,3}^{\prime,\prime\prime}$ are defined in the main text, note that for the high symmetry structure $J_{1,3}^{\prime} = J_{1,3}^{\prime\prime} = J_{1,3}$.  The crystal is oriented as in the Figure \ref{fig:distortion-summary} and the neighbours are defined by the vectors $\delta_{1-3}$ that connects the $i$-th \ce{Ni} atom with its first nearest neighbours. The relation between the exchange parameters and connecting vectors is summarized in the Table \ref{table:neighbors-parameters}, where $a = 3.895$~\AA\  is an in-plane lattice parameter and $\boldsymbol{\delta}_1 = (a, 0, 0)$, $\boldsymbol{\delta}_2 = (a/2, \sqrt{3}a/2, 0)$, $\boldsymbol{\delta}_3 = (-a/2, \sqrt{3}a/2, 0)$.

\begin{table}[H]
    \centering
    \begin{tabular}{|c|c|c|c|c|}
        \hline
        Parameters & $J_1^{\prime}$             & $J_1^{\prime\prime}$                                    & $J_3^{\prime}$              & $J_3^{\prime\prime}$                                    \\ \hline
        Neighbors  & $\pm\boldsymbol{\delta}_1$ & $\pm\boldsymbol{\delta}_2$, $\pm\boldsymbol{\delta}_3$  & $\pm2\boldsymbol{\delta}_1$ & $\pm\boldsymbol{\delta}_2$, $\pm2\boldsymbol{\delta}_3$ \\ \hline
    \end{tabular}
    \caption{Relation between the neighbours and exchange parameters}
    \label{table:neighbors-parameters}
\end{table}

The solution based on the approach from \cite{Toth2015}, which is correct in the case of the isotropic exchange parameters in the Hamiltonian, but should be treated carefully for the case with the full exchange matrix.
We solve the magnon problem for the conical spiral state that is defined by the spin spiral vector $\mathbf{q} = (q_x, q_y, 0)^T$ and rotation vector $\mathbf{e} = (0,0,1)^T$. Then the spin vectors can be written as follows:

\begin{equation}
    \mathbf{S}_i = \mathbf{R}_i\cdot\mathbf{R}\cdot\mathbf{S^{c}}_i,
\end{equation}
where $\mathbf{S^{c}}_i$ - is a set of collinear spins oriented along the direction of $z$ axis. Note, that although the energy of the spiral system does not depend on the choice of the plane for the rotation of the spiral, but the magnon's energy depend on it. Next, we applied linearized Holstein-Primakoff \cite{Holstein1940} expansion to this collinear state:

\begin{align}
    \mathbf{S}_i^{\boldsymbol{c}, x} &= \sqrt{\dfrac{S}{2}} (b_i + b_i^{\dagger}) \\
    \mathbf{S}_i^{\boldsymbol{c}, y} &= \dfrac{1}{i}\sqrt{\dfrac{S}{2}} (b_i - b_i^{\dagger}) \\
    \mathbf{S}_i^{\boldsymbol{c}, z} &= (S - b_i^{\dagger}b_i) \\
\end{align}

Rotation matrix $\mathbf{R}$ does not depend on the unit cell index $i$ and rotates the spins from the collinear stated oriented along the $z$ axis to the collinear state oriented along the $x$ axis:

\begin{equation}
    \mathbf{R} = 
    \begin{pmatrix}
         0 & 0 & 1 \\
         0 & 1 & 0 \\
        -1 & 0 & 0 \\
    \end{pmatrix}
\end{equation}

To simplify the notation we define two vectors $\mathbf{u} = (0,i,-1)^T$ and $\mathbf{v} = (1,0,0)^T$, then the array of spins oriented along the $x$ axis can be written as follows

\begin{equation}
    \mathbf{R}\cdot\mathbf{S^{c}}_i = \sqrt{\dfrac{S}{2}} \mathbf{\bar{u}} b_i
    + \sqrt{\dfrac{S}{2}} \mathbf{u} b_i^{\dagger}
    + \mathbf{v}(S - b_i^{\dagger}b_i),
\end{equation}
where $\mathbf{\bar{u}}$ denotes complex conjugate. Finally, we arrive at the spiral state by rotating each individual spin around the $\mathbf{\hat{n}}$ axis by the angle $\mathbf{q}\cdot\mathbf{r}_i$ ($\mathbf{r}_i$ is the position of the $i$~atom):

\begin{equation}
    \mathbf{R}_i = 
    \begin{pmatrix}
        \cos(\mathbf{q}\cdot\mathbf{r}_i) & -\sin(\mathbf{q}\cdot\mathbf{r}_i) & 0 \\
        \sin(\mathbf{q}\cdot\mathbf{r}_i) &  \cos(\mathbf{q}\cdot\mathbf{r}_i) & 0 \\
        0 & 0 & 1 \\ 
    \end{pmatrix}
\end{equation}

The full Hamiltonian is

\begin{multline}\label{Eq:hamiltonian}
    \mathcal{H} \approx \sum_{i} \sum_j\sum_{\delta = \pm\delta_j}
    \left(\sqrt{\dfrac{S}{2}} \mathbf{\bar{u}} b_i
    + \sqrt{\dfrac{S}{2}} \mathbf{u} b_i^{\dagger}
    + \mathbf{v}(S - b_i^{\dagger}b_i)\right)^T \cdot \\
    \cdot \mathbf{R}_i^T J_j \mathbf{R}_{i+\delta} \cdot \\
    \cdot \left(\sqrt{\dfrac{S}{2}} \mathbf{\bar{u}} b_{i+\delta}
    + \sqrt{\dfrac{S}{2}} \mathbf{u} b_{i+\delta}^{\dagger}
    + \mathbf{v}(S - b_{i+\delta}^{\dagger}b_{i+\delta})\right)
\end{multline}

In the case of isotropic exchange we can simplify the expression by defining

\begin{equation}
    \mathbf{J}_{j,\delta} = \mathbf{R}_i^T J_j \mathbf{R}_{i+\delta} = J_j \mathbf{R}_{\delta} =
    \begin{pmatrix}
        J_j\cos(\mathbf{q}\cdot\boldsymbol{\delta}) & -J_j\sin(\mathbf{q}\cdot\boldsymbol{\delta}) & 0 \\
        J_j\sin(\mathbf{q}\cdot\boldsymbol{\delta}) &  J_j\cos(\mathbf{q}\cdot\boldsymbol{\delta}) & 0 \\
        0 & 0 & J \\ 
    \end{pmatrix}
\end{equation}
which does not depend on the lattice index $i$. Expanding the parenthesis and cutting the expression at the level of linearised spin wave theory (LSWT) (i.e. by keeping only the terms of the order of $1/S$ and higher and omitting one-operator terms) we get

\begin{multline}
    \mathcal{H} \approx S^2\sum_{i} \sum_j\sum_{\delta = \pm\delta_j}\Biggl[ \mathbf{v}^T\mathbf{J}_{j,\delta}\mathbf{v}\\
    + \dfrac{1}{2S}\Bigl(
      \mathbf{\bar{u}}^T \mathbf{J}_{j,\delta} \mathbf{\bar{u}}      b_i           b_{i+\delta}
    + \mathbf{\bar{u}}^T \mathbf{J}_{j,\delta} \mathbf{u}            b_i           b_{i+\delta}^{\dagger}
    + \mathbf{u}^T       \mathbf{J}_{j,\delta} \mathbf{\bar{u}}      b_i^{\dagger} b_{i+\delta}
    + \mathbf{u}^T       \mathbf{J}_{j,\delta} \mathbf{u}            b_i^{\dagger} b_{i+\delta}^{\dagger} \\
    - 2\mathbf{v}^T \mathbf{J}_{j,\delta} \mathbf{v} b_i^{\dagger} b_i
    - 2\mathbf{v}^T \mathbf{J}_{j,\delta} \mathbf{v} b_{i+\delta}^{\dagger} b_{i+\delta}
    \Bigr)
    \Biggr]
\end{multline}

Next, we rewrite the last term of the sum using the commutator $[b_i, b_i^{\dagger}] = 1$, to show explicitly that the Hamiltonian is hermitian:

\begin{equation}
    -2\mathbf{v}^T \mathbf{J}_{j,\delta} \mathbf{v} b_{i+\delta}^{\dagger} b_{i+\delta} = -2\mathbf{v}^T \mathbf{J}_{j,\delta} \mathbf{v} b_{i+\delta} b_{i+\delta}^{\dagger} + 2\mathbf{v}^T \mathbf{J}_{j,\delta} \mathbf{v}
\end{equation}

Finally, the Hamiltonian consist of three parts

\begin{equation}
    \mathcal{H} = \mathcal{H}^{0, classic} + \mathcal{H}^{0, correction} + \mathcal{H}^{LSWT} + S\cdot\mathcal{O}\left(\dfrac{1}{S}\right),
\end{equation}
where the ground state energy reads as

\begin{equation}
    E = \mathcal{H}^{0, classic} + \mathcal{H}^{0, correction} = N S^2(1 + \dfrac{1}{S})\sum_j\sum_{\delta = \pm\delta_j} \mathbf{v}^T \mathbf{J}_{j,\delta} \mathbf{v} = 2N S^2(1 + \dfrac{1}{S})\sum_j J_j \cos(\mathbf{q}\cdot\boldsymbol{\delta_j}),
\end{equation}
where $N$ is the amount of unit cells in the crystal. Note, that quantum correction to the classical ground state energy adds a multiplier to the value of the ground state energy, but does not change its character. Now, we solve the LSWT part of the Hamiltonian.

\begin{multline}
    \mathcal{H}^{LSWT} = S\sum_{i} \sum_j\sum_{\delta = \pm\delta_j} \Biggl[
      \dfrac{}{2}\mathbf{\bar{u}}^T \mathbf{J}_{j,\delta} \mathbf{\bar{u}}      b_i           b_{i+\delta}
    + \dfrac{}{2}\mathbf{\bar{u}}^T \mathbf{J}_{j,\delta} \mathbf{u}            b_i           b_{i+\delta}^{\dagger}
    + \dfrac{}{2}\mathbf{u}^T       \mathbf{J}_{j,\delta} \mathbf{\bar{u}}      b_i^{\dagger} b_{i+\delta}
    + \dfrac{}{2}\mathbf{u}^T       \mathbf{J}_{j,\delta} \mathbf{u}            b_i^{\dagger} b_{i+\delta}^{\dagger} \\
    - \mathbf{v}^T \mathbf{J}_{j,\delta} \mathbf{v} b_i^{\dagger} b_i
    - \mathbf{v}^T \mathbf{J}_{j,\delta} \mathbf{v} b_{i+\delta} b_{i+\delta}^{\dagger}
    \Biggr]
\end{multline}

First, we explicitly compute matrix products

\begin{multline}
    \mathcal{H}^{LSWT} = S\sum_{i} \sum_j J_j \sum_{\delta = \pm\delta_j} \Biggl[
      \dfrac{1 - \cos(\mathbf{q}\cdot\boldsymbol{\delta})}{2}(b_i b_{i+\delta} + b_i^{\dagger} b_{i+\delta}^{\dagger}) + \\   
    + \dfrac{1 + \cos(\mathbf{q}\cdot\boldsymbol{\delta})}{2}(b_i b_{i+\delta}^{\dagger} + b_i^{\dagger} b_{i+\delta}) - \\
    - \cos(\mathbf{q}\cdot\boldsymbol{\delta}) (b_i^{\dagger} b_i + b_{i+\delta} b_{i+\delta}^{\dagger})
    \Biggr]
\end{multline}

Second, we apply Fourier transformation

\begin{align}
    b_i = \dfrac{1}{\sqrt{N}}\sum_k e^{i\mathbf{k}\cdot\mathbf{r}_i} b_k \\
    b_i^{\dagger} = \dfrac{1}{\sqrt{N}}\sum_k e^{-i\mathbf{k}\cdot\mathbf{r}_i} b_k^{\dagger}
\end{align}

\begin{multline}
    \mathcal{H}^{LSWT} = S\sum_{i} \sum_j J_j \sum_{\delta = \pm\delta_j} \sum_k \sum_{k^{\prime}} \dfrac{1}{N} \Biggl[
      \dfrac{1 - \cos(\mathbf{q}\cdot\boldsymbol{\delta})}{2}
      \left(e^{i\mathbf{k}\cdot\mathbf{r}_i}e^{i\mathbf{k^{\prime}}\cdot(\mathbf{r}_i+\boldsymbol{\delta})} b_k b_{k^{\prime}} + e^{-i\mathbf{k}\cdot\mathbf{r}_i}e^{-i\mathbf{k^{\prime}}\cdot(\mathbf{r}_i+\boldsymbol{\delta})} b_k^{\dagger} b_{k^{\prime}}^{\dagger}\right) +  \\   
      \phantom{\mathcal{H}^{LSWT} = JS\sum_{i} \sum_{\pm\delta} \sum_k \sum_{k^{\prime}} \dfrac{1}{N} \Biggl[}
    + \dfrac{1 + \cos(\mathbf{q}\cdot\boldsymbol{\delta})}{2}
      \left(e^{i\mathbf{k}\cdot\mathbf{r}_i}e^{-i\mathbf{k^{\prime}}\cdot(\mathbf{r}_i+\boldsymbol{\delta})} b_k b_{k^{\prime}}^{\dagger} + e^{-i\mathbf{k}\cdot\mathbf{r}_i}e^{i\mathbf{k^{\prime}}\cdot(\mathbf{r}_i+\boldsymbol{\delta})} b_k^{\dagger} b_{k^{\prime}}\right) - \\
    - \cos(\mathbf{q}\cdot\boldsymbol{\delta})
      \left(e^{-i\mathbf{k}\cdot\mathbf{r}_i}e^{i\mathbf{k^{\prime}}\cdot\mathbf{r}_i} b_k^{\dagger} b_{k^{\prime}} + e^{i\mathbf{k}\cdot\mathbf{r}_i}e^{-i\mathbf{k^{\prime}}\cdot\mathbf{r}_i} b_k b_{k^{\prime}}^{\dagger}\right)
    \Biggr]
\end{multline}
and use orthogonality condition $\frac{1}{N}\sum_ie^{i(\mathbf{k}-\mathbf{k}^{\prime})\mathbf{r}_i} = \delta_{k,k^{\prime}}$

\begin{multline}
    \mathcal{H}^{LSWT} = S \sum_j J_j \sum_{\delta = \pm\delta_j} \sum_k \Biggl[
      \dfrac{1 - \cos(\mathbf{q}\cdot\boldsymbol{\delta})}{2}
      \left(e^{-i\mathbf{k}\cdot\boldsymbol{\delta}} b_k b_{-k} + e^{i\mathbf{k}\cdot\boldsymbol{\delta}} b_k^{\dagger} b_{-k}^{\dagger}\right) + \\   
    + \dfrac{1 + \cos(\mathbf{q}\cdot\boldsymbol{\delta})}{2}
      \left(e^{-i\mathbf{k}\cdot\boldsymbol{\delta}} b_k b_{k}^{\dagger} + e^{i\mathbf{k}\cdot\boldsymbol{\delta}} b_k^{\dagger} b_{k}\right) - \\
    - \cos(\mathbf{q}\cdot\boldsymbol{\delta})
      \left(b_k^{\dagger} b_{k} + b_k b_{k}^{\dagger}\right)
    \Biggr]
\end{multline}

and expanding the sum over $\pm\delta_j$

\begin{multline}
    \mathcal{H}^{LSWT} = S \sum_k \sum_j J_j\Biggl[
      \left(1 - \cos(\mathbf{q}\cdot\boldsymbol{\delta}_j)\right)\cos(\mathbf{k}\cdot\boldsymbol{\delta}_j)
      \left(b_k b_{-k} + b_k^{\dagger} b_{-k}^{\dagger}\right) + \\   
    + \left(1 + \cos(\mathbf{q}\cdot\boldsymbol{\delta}_j)\right)\cos(\mathbf{k}\cdot\boldsymbol{\delta}_j)
      \left(b_k b_{k}^{\dagger} + b_k^{\dagger} b_{k}\right) - \\
    - 2\cos(\mathbf{q}\cdot\boldsymbol{\delta}_j)
      \left(b_k^{\dagger} b_{k} + b_k b_{k}^{\dagger}\right)
    \Biggr]
\end{multline}

Finally we define $\mathbf{X}_k = (b_k, b_{-k}^\dagger)^T$ and write the Hamiltonian in a quadratic form (note: $[b_{-k},b_k] = [b_{-k}^{\dagger}, b_{k}^{\dagger}] = 0$ and $b_kb_k^{\dagger} = b_{-k}b_{-k}^{\dagger}$)

\begin{equation}
    \mathcal{H}^{LSWT} = S \sum_k \mathbf{X}_k^{\dagger} \mathbf{h}(\mathbf{k}) \mathbf{X}_k^{\dagger},
\end{equation}
where the Hamiltonian matrix is Hermitian and reads as follows

\begin{equation}
    \mathbf{h}(\mathbf{k}) = 
    \begin{pmatrix}
        A(\boldsymbol{k}) & B(\boldsymbol{k}) \\ 
        B(\boldsymbol{k}) & A(\boldsymbol{k}) \\
    \end{pmatrix}
\end{equation}
where 

\begin{equation}
    A(\boldsymbol{k}) = \sum_j J_j \biggl(\bigl[1 + \cos(\mathbf{q}\cdot\boldsymbol{\delta}_j)\bigr]\cdot\cos(\mathbf{k}\cdot\boldsymbol{\delta}_j) - 2\cos(\mathbf{q}\cdot\boldsymbol{\delta}_j)\biggr),
\end{equation}
\begin{equation}
    B(\boldsymbol{k}) = \sum_j J_j \bigl[1 - \cos(\mathbf{q}\cdot\boldsymbol{\delta}_j)\bigr]\cdot\cos(\mathbf{k}\cdot\boldsymbol{\delta}_j)
\end{equation}

The only remaining part left is the diagonalization of the bosonic Hamiltonian. We follow approach taken by White et al. \cite{White1965} and find an expression for the diagonalized Hamiltonian: 

\begin{equation}
    \mathcal{H}^{LSWT} = S \sum_k \mathbf{Y}_k^{\dagger} \boldsymbol{\mathcal{\varepsilon}}(\mathbf{k}) \mathbf{Y}_k^{\dagger},
\end{equation}
where $\mathbf{Y}_k = (\beta_k, \beta_k^{\dagger})^T$ are the new bosonic operators, that obey the same commutation relation: $[\beta_k\beta_k^{\dagger}] = 1$ and matrix $\boldsymbol{\mathcal{\varepsilon}}(\mathbf{k})$ is diagonal:

\begin{equation}
    \boldsymbol{\mathcal{\varepsilon}}(\mathbf{k})
    =
    \begin{pmatrix}
        \sqrt{A^2(\boldsymbol{k}) - B^2(\boldsymbol{k})} & 0 \\
        0 & \sqrt{A^2(\boldsymbol{k}) - B^2(\boldsymbol{k})} \\
    \end{pmatrix}
\end{equation}

Now turning back to the scalar form and renaming $-k -> k$ in the second term we get (note, that $A(-\boldsymbol{k}) = A(\boldsymbol{k})$ and $B(-\boldsymbol{k}) = B(\boldsymbol{k})$):

\begin{align}
    \mathcal{H}^{LSWT}
    &=
    S \sum_k
    \sqrt{A^2(\boldsymbol{k}) - B^2(\boldsymbol{k})}
    \bigl[
      \beta_k^{\dagger}\beta_k + \beta_k\beta_k^{\dagger}
    \bigr]
    =\\\phantom{\mathcal{H}^{LSWT}}&=
    S \sum_k
    \sqrt{A^2(\boldsymbol{k}) - B^2(\boldsymbol{k})}
    \bigl[
      2\beta_k^{\dagger}\beta_k + 1
    \bigr]
    =\\\phantom{\mathcal{H}^{LSWT}}&=
    \sum_k
    \hbar\omega(\boldsymbol{k})
    \biggl[
      \beta_k^{\dagger}\beta_k + \dfrac{1}{2}
    \biggr]
\end{align}

where eigenfrequencies are given by:

\begin{equation}
    \hbar\omega(\mathbf{k}) = 2S\sqrt{A^2(\boldsymbol{k}) - B^2(\boldsymbol{k})}
\end{equation}

\begin{figure}[H]
     \centering
     \begin{subfigure}[c]{0.3\textwidth}
         \centering
         \caption{}
         \includegraphics[width=\textwidth]{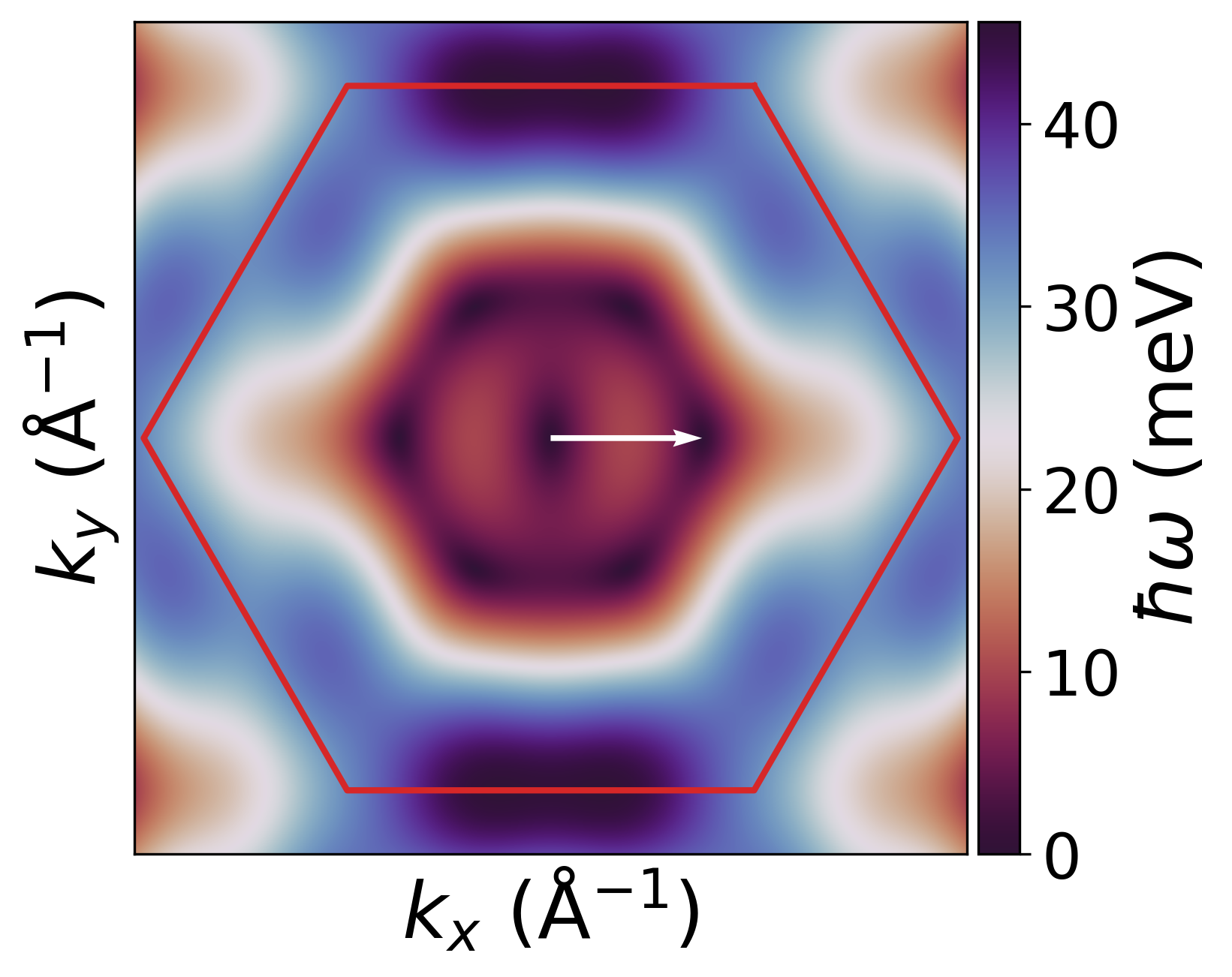}\label{Fig:magnons_dispersion-qK1}
     \end{subfigure}
     \hfill
     \begin{subfigure}[c]{0.3\textwidth}
         \centering
         \caption{}
         \includegraphics[width=\textwidth]{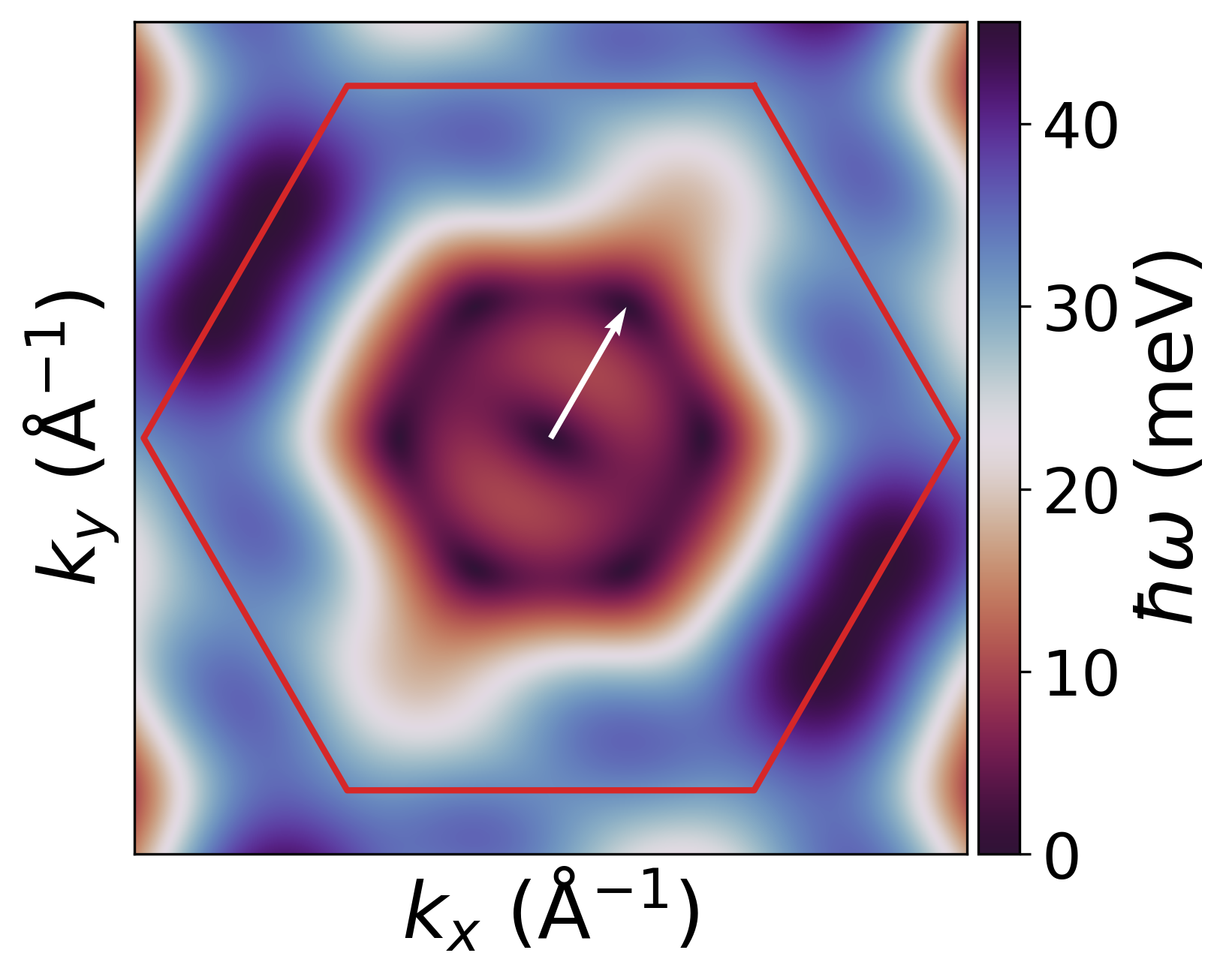}\label{Fig:magnons_dispersion-qK}
     \end{subfigure}
     \hfill
     \begin{subfigure}[c]{0.3\textwidth}
         \centering
         \caption{}
         \includegraphics[width=\textwidth]{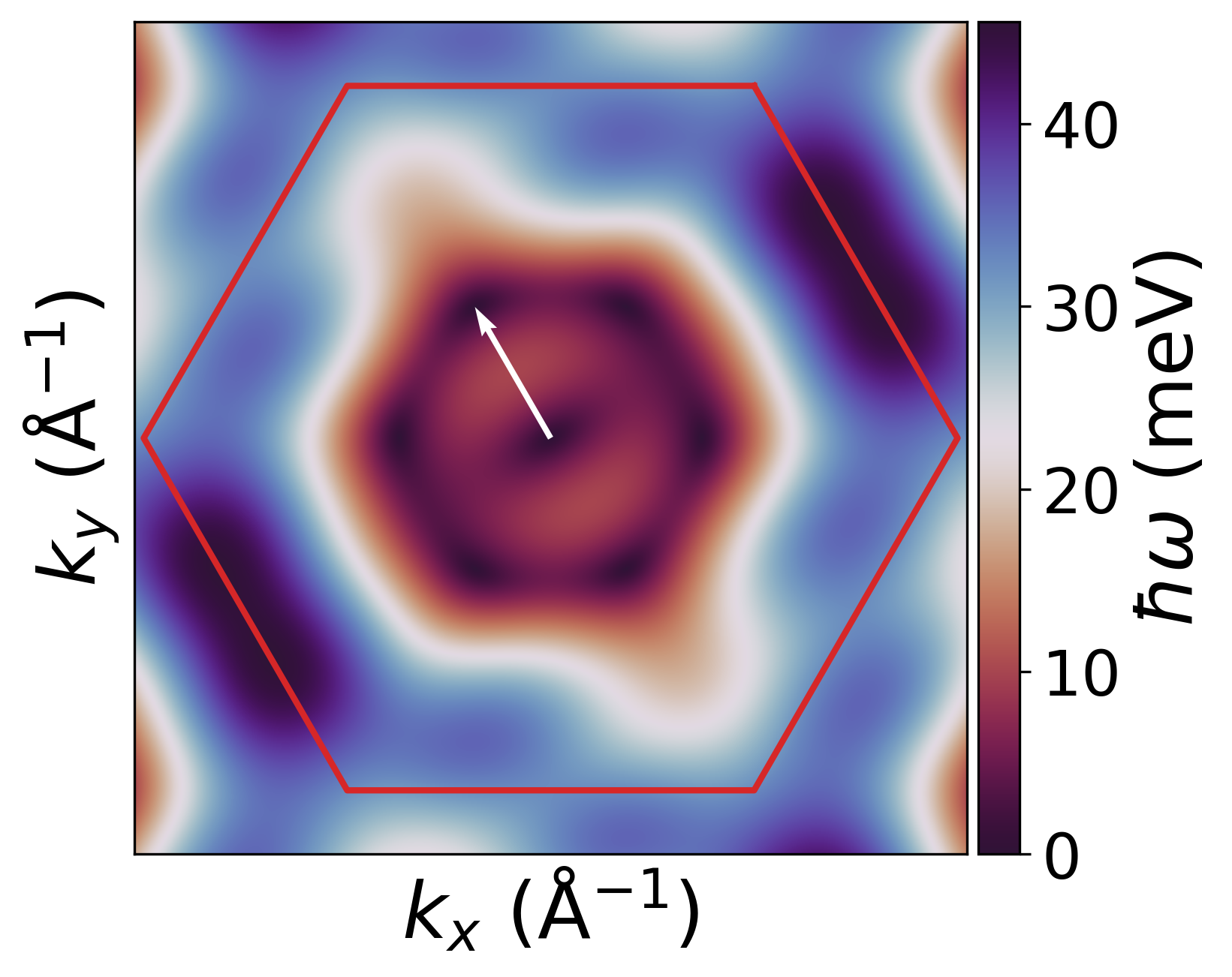}\label{Fig:magnons_dispersion-qK2}
     \end{subfigure}
        \caption{Magnon dispersion for the C$_6$ symmetric undistorted structure with the spiral vector along (a) $\Gamma$K$_1$, (b) $\Gamma$K, (c) $\Gamma$K$_2$ directions. The $q$ vector of spin spiral is depicted as white arrow. First Brillouin zone is indicated by red hexagon.}
        \label{Fig:magnons_dispersion_undist}
\end{figure}

In the Figure~\ref{Fig:magnons_dispersion_undist} magnon energies of the first Brillouin zone are displayed for three degenerate spiral states. 

\begin{figure}[H]
     \centering
     \begin{subfigure}[c]{0.3\textwidth}
         \centering
         \caption{}
         \includegraphics[width=\textwidth]{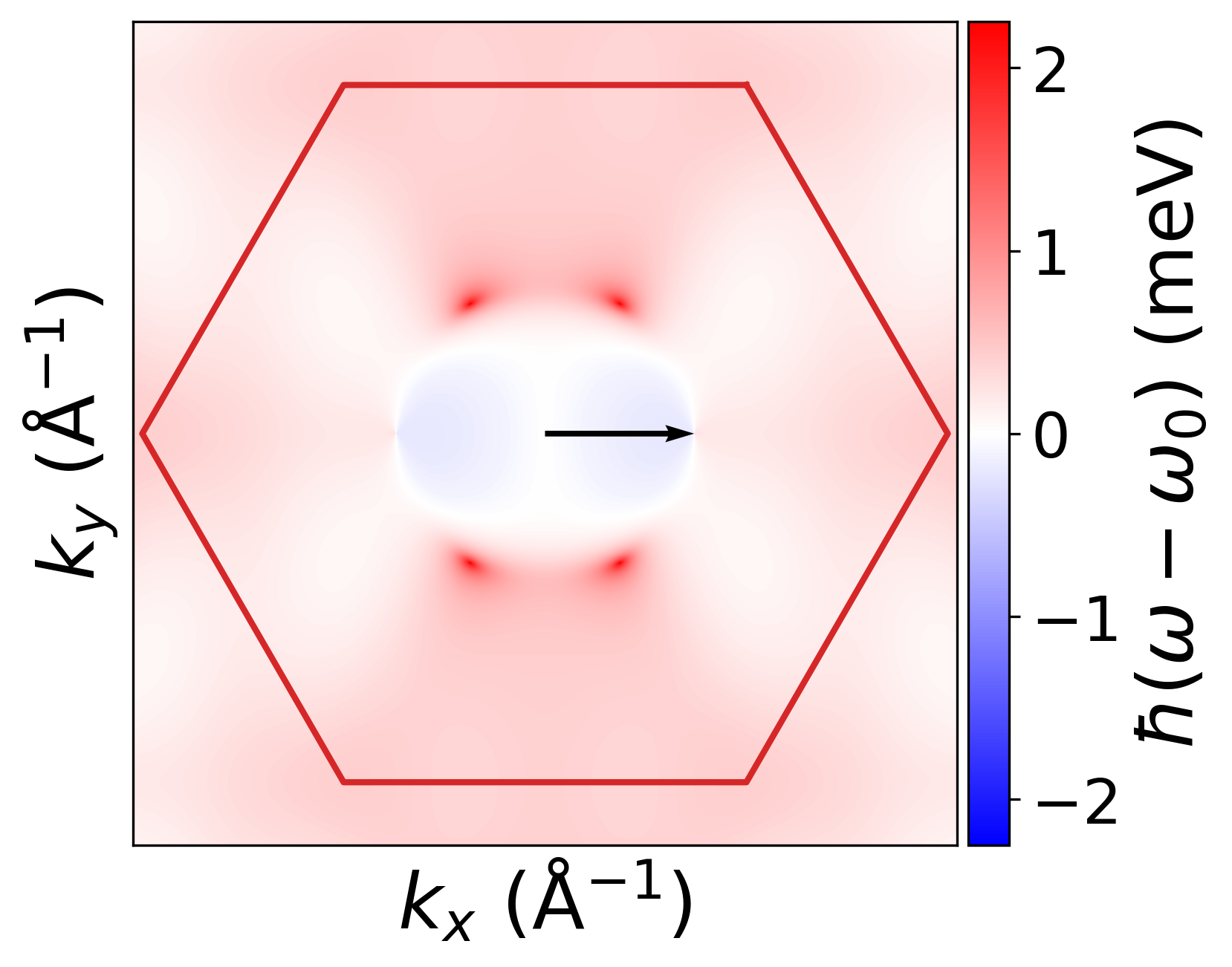}\label{Fig:magnons_dispersion-e0-relative}
     \end{subfigure}
     \hfill
     \begin{subfigure}[c]{0.3\textwidth}
         \centering
         \caption{}
         \includegraphics[width=\textwidth]{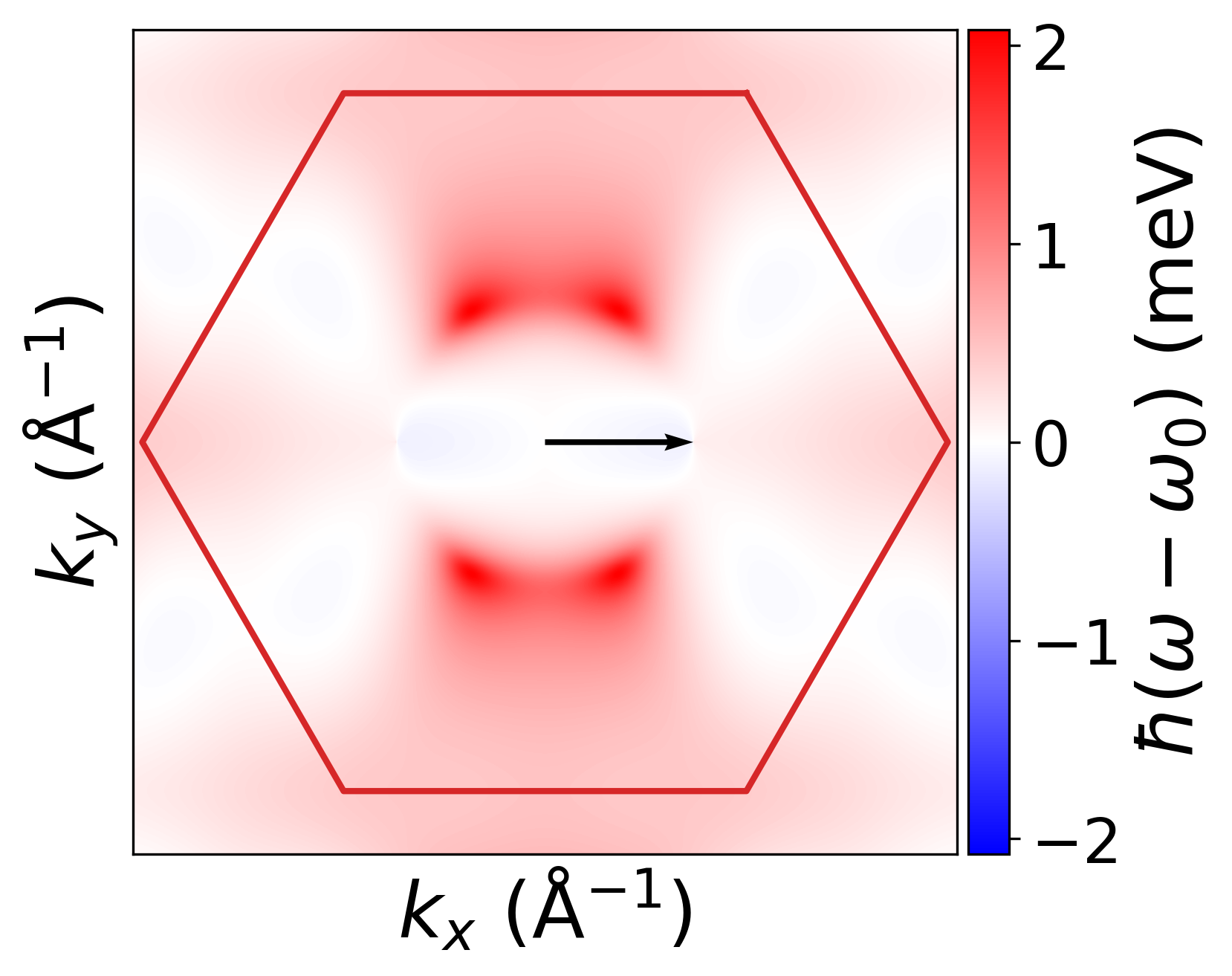}\label{Fig:magnons_dispersion-e50-relative}
     \end{subfigure}
     \hfill
     \begin{subfigure}[c]{0.3\textwidth}
         \centering
         \caption{}
         \includegraphics[width=\textwidth]{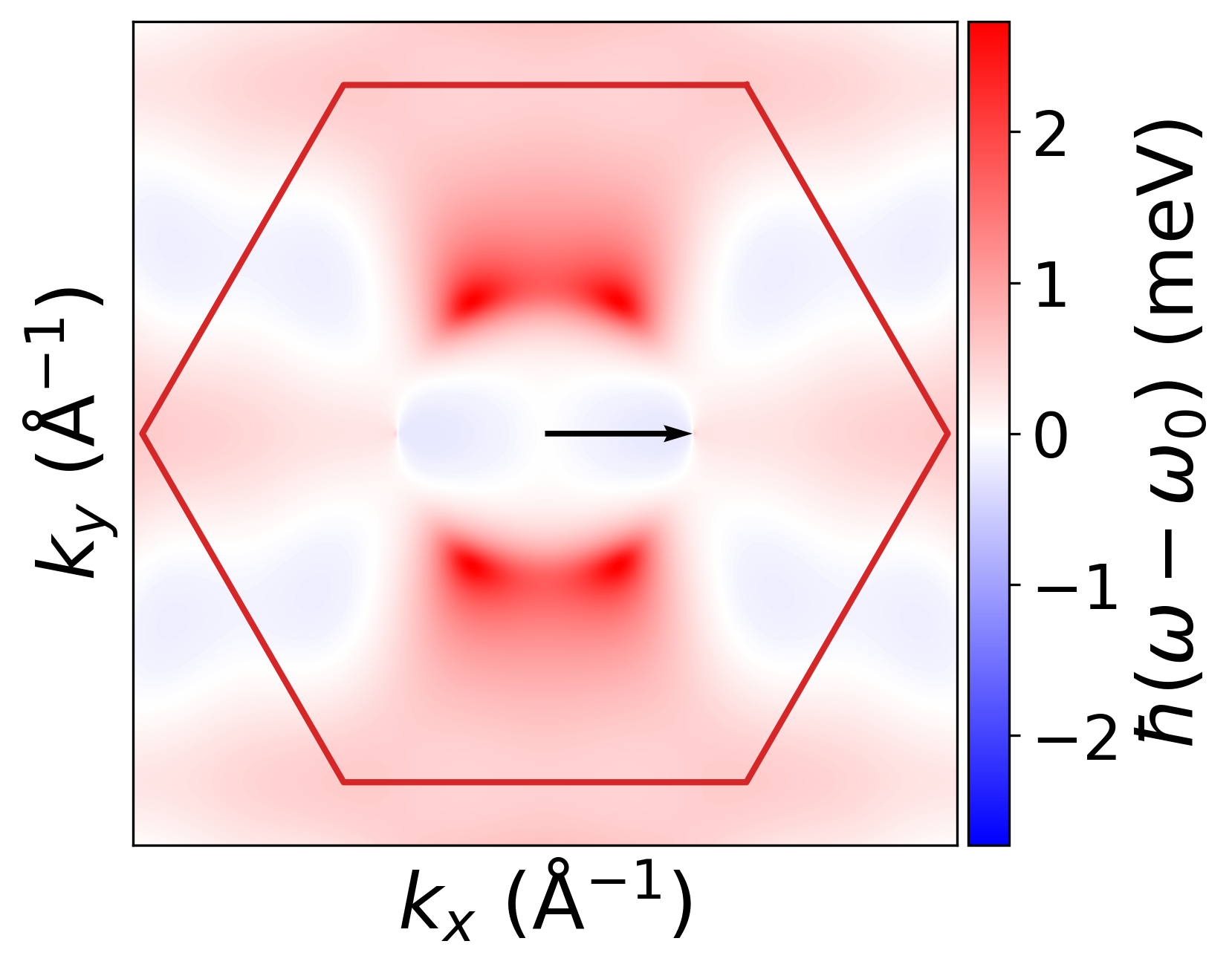}\label{Fig:magnons_dispersion-e100-relative}
     \end{subfigure}
        \caption{Differences of magnon energies for the C$_2$ symmetric structure under the applied values of electric field (a) $0$~mV/\AA, (b) $50$~mV/\AA, (c) $100$~mV/\AA. The $\mathbf{q}$ vector of spin spiral is depicted as black arrow. First Brillouin zone is indicated by red hexagon. $\hbar\omega_0$ is the energy of magnons in the (a) C$_6$, (b-c) C$_2$ symmetric structure with the spin spiral vector oriented along the $\Gamma$K$_1$ direction.}
        \label{Fig:magnons_dispersion_dist_delta}
\end{figure}

In the Figure~\ref{Fig:magnons_dispersion_dist_delta} the relative change of the magnon energy with respect to the undistorted structure is shown. One can see that the directions along the $\Gamma$K and $\Gamma$K$_2$ are equivalent from the perspective of the spin wave exitations.

\begin{figure}[H]
     \centering
     \begin{subfigure}[c]{0.47\textwidth}
         \centering
         \caption{}
         \includegraphics[width=\textwidth]{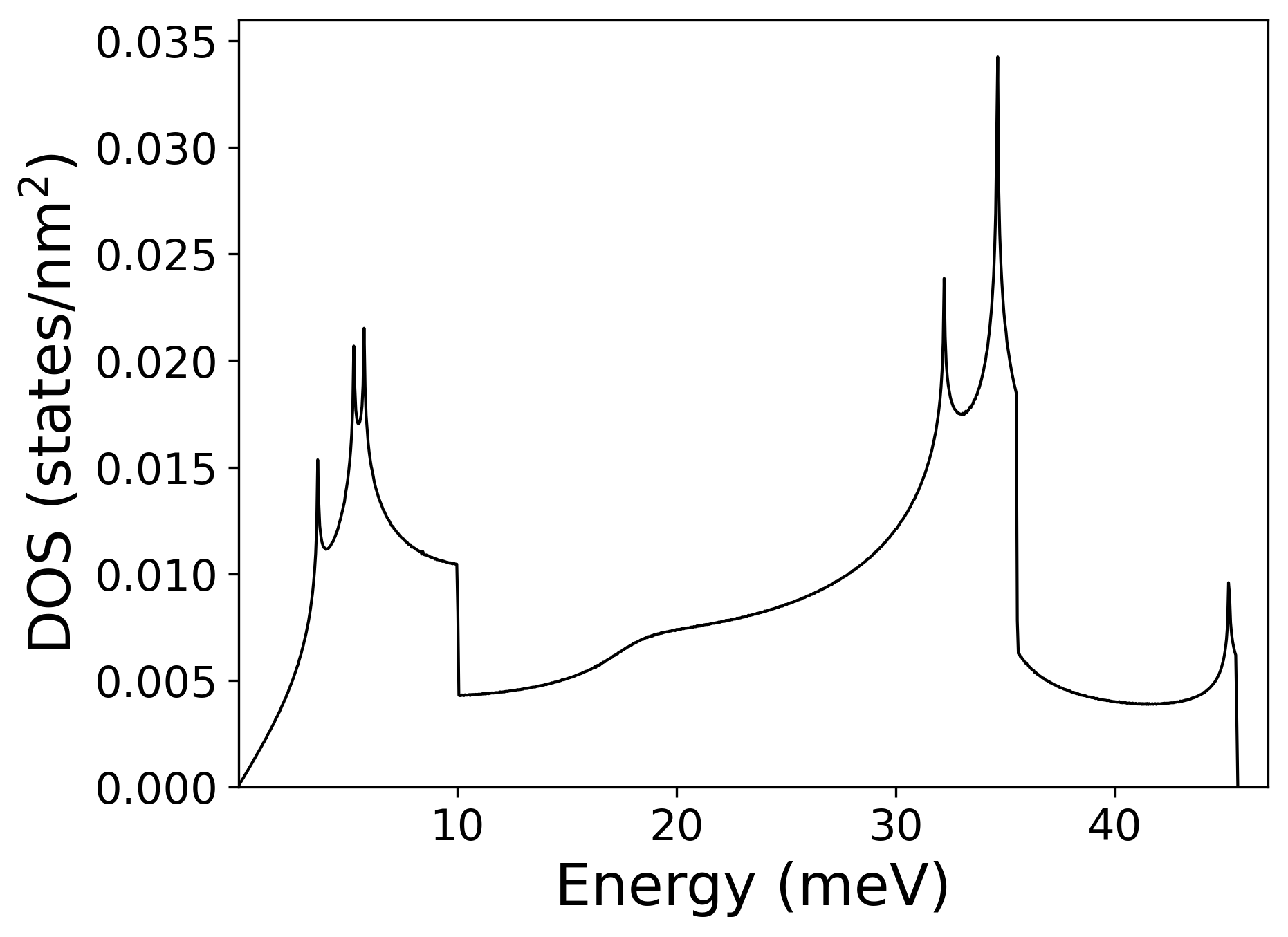}\label{Fig:magnons_dos-undist-qK1}
     \end{subfigure}
     \hfill
     \begin{subfigure}[c]{0.47\textwidth}
         \centering
         \caption{}
         \includegraphics[width=\textwidth]{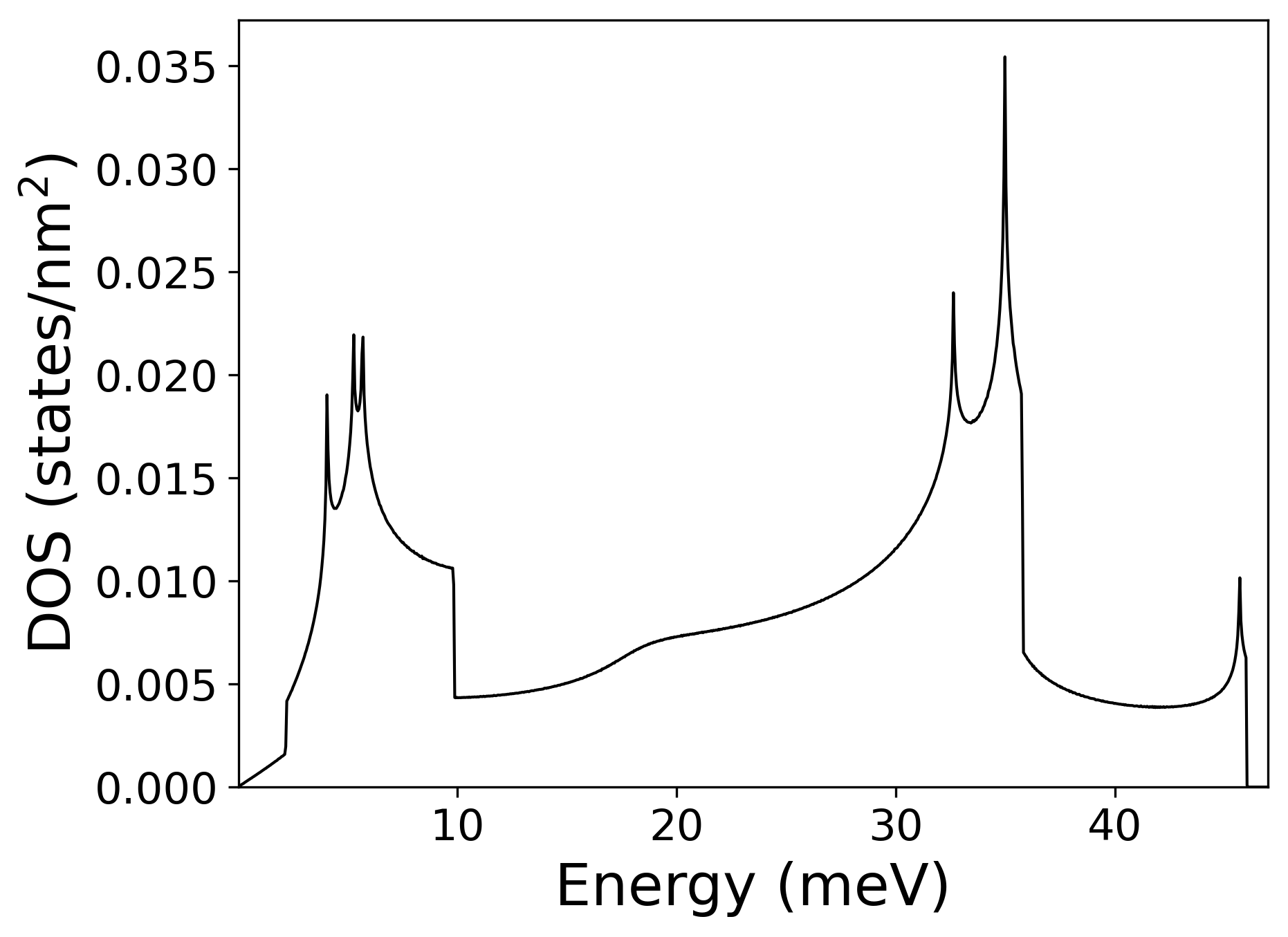}\label{Fig:magnons_dos-e0}
     \end{subfigure}
     \hfill
     \begin{subfigure}[c]{0.47\textwidth}
         \centering
         \caption{}
         \includegraphics[width=\textwidth]{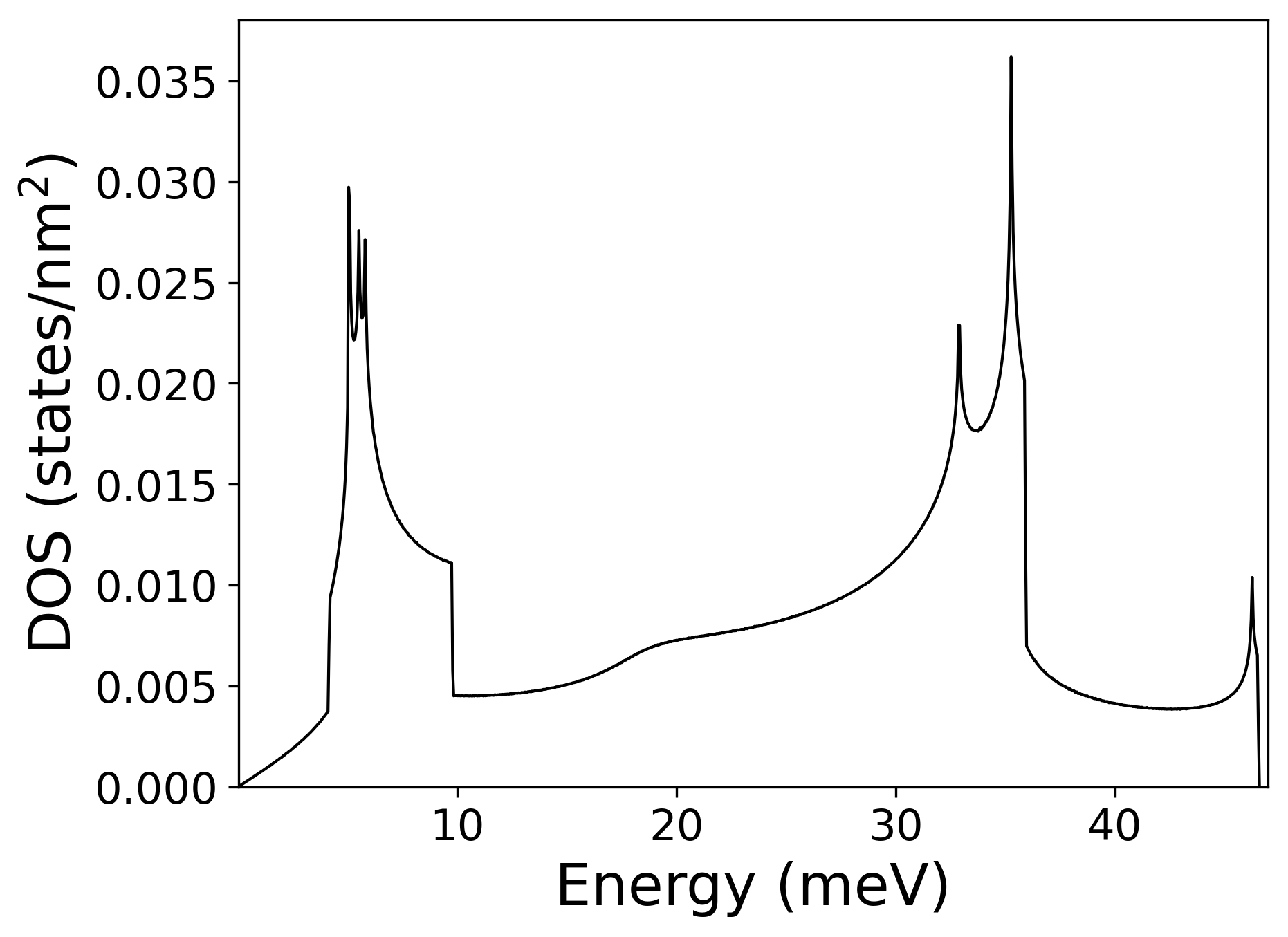}\label{Fig:magnons_dos-e50}
     \end{subfigure}
     \hfill
     \begin{subfigure}[c]{0.47\textwidth}
         \centering
         \caption{}
         \includegraphics[width=\textwidth]{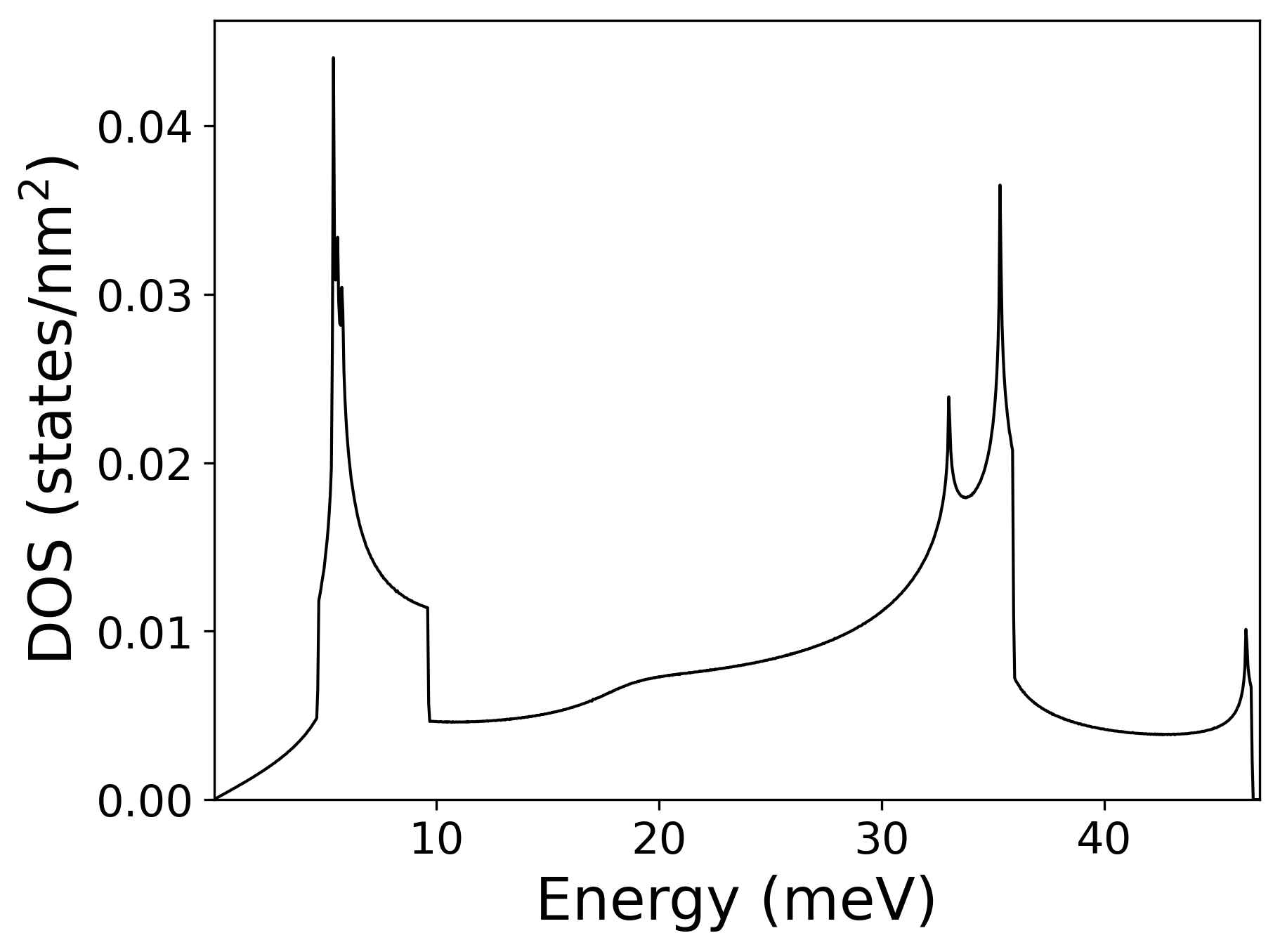}\label{Fig:magnons_dos-e100}
     \end{subfigure}
        \caption{Magnon density of states computed for three structures: (a) Undistorted structure (C$_6$), (b) Distorted structure without applied electric field, (c) Distorted structure (C$_2$) with applied electric field of $E = 50$~mV/\AA, (d) Distorted structure (C$_2$) with applied electric field of $E = 100$~mV/\AA.}
        \label{Fig:magnons_dos}
\end{figure}

Density of states (Figure~\ref{Fig:magnons_dos}) was computed numerically taking dense sampling of the first Brillouin zone.

\bibliography{biblio}